\documentstyle[12pt,cite]{article}

\newcommand{\pl}{\partial}
\newcommand{\be}{\begin{equation}}\newcommand{\ee}{\end{equation}}
\newcommand{\bea}{\begin{eqnarray}}\newcommand{\eea}{\end{eqnarray}}
\newcommand{\nn}{\nonumber}\newcommand{\p}[1]{(\ref{#1})}

\def\theequation{\arabic{section}.\arabic{equation}}
\begin{document}
\thispagestyle{empty}
\begin{flushright}ENSLAPP-L-554\\
JINR-E2-95-424 \\
hep-th/9510033 \\
October 1995
\end{flushright}
\vskip 1.0truecm
\begin{center}{\bf\Large
$N=4$ super KdV hierarchy in $N=4$ and $N=2$ superspaces}
\end{center}
 \vskip 1.0truecm
\centerline{\bf F. Delduc${}^{(a)}$, E. Ivanov${}^{(b)}$ and S. Krivonos
${}^{(b)}$}
\vskip 1.0truecm
\centerline{${}^{(a)}$ \it Lab. de Phys. Th\'eor. ENSLAPP, ENS Lyon}
\centerline{\it 46 All\'ee d'Italie, 69364 Lyon, France}
\vskip5mm
\centerline{${}^{(b)}$\it Bogoliubov Laboratory of Theoretical
Physics, JINR,}
\centerline{\it Dubna, 141 980 Moscow region, Russia}
\vskip 1.0truecm  \nopagebreak

\begin{abstract}
We present the results of further analysis of the integrability
properties of the $N=4$
supersymmetric KdV equation deduced earlier by
two of us (F.D. \& E.I., Phys. Lett. B 309 (1993) 312) as a hamiltonian
flow
on $N=4$ $SU(2)$ superconformal algebra in the harmonic $N=4$ superspace.
To make this equation and the relevant
hamiltonian structures more tractable, we reformulate it
in the ordinary $N=4$ and further in $N=2$
superspaces. In $N=2$ superspace it is represented by a coupled
system of evolution equations for a general $N=2$ superfield and two
chiral and antichiral superfields, and involves two independent real
parameters, $a$ and $b$. We construct a few first bosonic conserved
charges in involution, of dimensions from 1 to 6, and show that they
exist only for the following choices of the parameters:
(i) $a= 4, \;b=0$; (ii) $a= -2,\; b= -6$; (iii) $a= -2, \;b= 6$.
The same values are needed for the relevant evolution equations,
including $N=4$ KdV itself, to be bi-hamiltonian. We demonstrate that
the above three options are related via $SU(2)$ transformations
and actually amount to the $SU(2)$ covariant integrability condition
found in the harmonic superspace approach. Our results provide a
strong evidence that the unique $N=4$ $SU(2)$ super KdV hierarchy exists.
Upon reduction to $N=2$ KdV, the above three possibilities cease
to be equivalent. They give rise to  the $a=4$ and $a=-2$ $N=2$ KdV
hierarchies, which thus prove to be different truncations of the
single $N=4$ $SU(2)$ KdV one.

\end{abstract}
\setcounter{page}0
\newpage

\section{Introduction}

The Korteweg-de Vries (KdV) hierarchy and its supersymmetric extensions
were the subject of many studies for the last several years. Besides
supplying nice examples of integrable systems,
they bear a deep relation to conformal field theory, $2D$ gravity,
matrix models, etc.
One of the remarkable properties of these systems is that they
are related, via the second hamiltonian structure, to the classical
(super)conformal algebras: Virasoro algebra in the bosonic case and
$N\geq 1$ superconformal ones in the case of $N\geq 1$ superextended
hierarchies \cite{Mag}-\cite{DI}. Generalized KdV type systems
related to
$W_n$ algebras and their supersymmetric extensions  also received
a great deal of attention (see, e.g., Ref. 15 and references therein).

Up to now, supersymmetric KdV hierarchies have been constructed
for $N=1,2,3$ and $4$, based on the above mentioned relation to
superconformal algebras \cite{MR}-\cite{DI}.
An interesting peculiarity
is that, beginning with $N=2$, the supersymmetric KdV equations turn
out to be integrable (give rise to the whole hierarchy or,
in other words, have an infinite number of conservation laws
in involution) only for special
choices of the parameters in the hamiltonian. There exist only three
integrable $N=2$ KdV hierarchies: the $a=4$, $a=-2$ and $a=1$
ones \cite{{Mat1},{Mat2},{Pop}}, with $a$ a parameter entering into
the $N=2$ KdV hamiltonian, despite the fact that for any value of $a$
the related $N=2$ super KdV possesses $N=2$ SCA as the second
hamiltonian structure.
The generalized $N=2$ KdV system associated with $N=2$ $W_3$ algebra
(``$N=2$ super Boussinesq hierarchy'') has similar properties as
established in Refs. 16 and 15.
For the $N=3$ super KdV equation associated with $N=3$ SCA the
requirement of
integrability also strictly fixes the value of a free parameter in the
hamiltonian \cite{Yung}, though the existence of
the whole hierarchy in this case has not yet been proven
(the Lax pair representation has not been found).
Only a few higher order conservation laws in involution
have been constructed. Nevertheless the existence of such quantities is
highly non-trivial and provides strong evidence in favour of the
integrability of the associated $N=3$ super KdV.

Another higher $N$ extension of $N=2$ super KdV, the $N=4$ one, has been
constructed in the article of two of us \cite{DI}. We proceeded from
the $N=4$ $SU(2)$ ("small") superconformal algebra \cite{a9} as the second
hamiltonian structure. This extension is, in a sense, more economic than
the $N=3$ one, because $N=4$ $SU(2)$ SCA by its
component currents content is a natural generalization of $N=2$ SCA.
Like $N=2$ SCA, it contains only currents with canonical dimensions:
a dimension 2
conformal stress tensor, four dimension $3/2$ fermionic currents and
three
dimension 1 affine $su(2)$ currents. $N=3$ SCA includes an extra
current with a subcanonical dimension $1/2$ \cite{a9}. Both $N=4$ $SU(2)$
and $N=2$ SCAs belong to the family of $u(N)$ Knizhnik-Bershadsky
superconformal algebras (which are nonlinear in general, starting
with $N=3$).

In our construction \cite{DI} we used the formalism of
$N=4$, $1D$ harmonic superspace (HSS)
as the most natural one for representing $N=4$ $SU(2)$ SCA  in a
manifestly supersymmetric form. We found that
the general superfield $N=4$ KdV hamiltonian, $H_3$, consists of two
pieces. One
is an integral over the whole $N=4$ HSS and the
second is an integral over an analytic subspace of this HSS, containing
half the number of odd coordinates. This second piece involves
a set of $SU(2)$ breaking constants which are naturally
combined into a symmetric
rank 4 $SU(2)$ spinor $c^{ilkj}$ (symmetric traceless rank 2 tensor).
We did not construct a Lax pair for the $N=4$ KdV equation, but
instead addressed the question of the existence of higher order conserved
quantities, like in the
$N=3$ KdV case \cite{Yung}.
We found that such quantities  exist and, hence, that
$N=4$ KdV can lead to
an integrable hierarchy, provided (i) the $SU(2)$ breaking tensor
is expressed as a square of some constant real $SU(2)$ vector
$a^{ij} = a^{ji}$, $(i,j =1,2)$, and (ii) the norm of the latter is
proportional to the reciprocal of the level of the
affine $su(2)$ subalgebra of $N=4$ $SU(2)$ SCA
\bea
c^{ijkl} &=& {1\over 3} \left( a^{ij} a^{kl} + a^{ik}a^{jl} +
a^{il}a^{jk}
\right)  \label{consa1} \\
|a|^{2} &\equiv & - a^{ij}a_{ij} = {20\over k}\;. \label{consa}
\eea
We also showed that under these restrictions the $N=4$ KdV equation
is bi-hamiltonian, i.e. it possesses a first hamiltonian structure,
the relevant
hamiltonian being the dimension 4 conserved charge $H_4$ (next
in dimension to $H_3$). We considered a reduction to the $N=2$ case and
found that, under a certain embedding of $U(1)$ subalgebra in $SU(2)$,
the $a=4$ integrable version \cite{Mat1} of $N=2$ KdV comes out.

In Ref. 14 we limited ourselves to the construction of the dimension 4
higher order conserved charge. On the other hand, it is known that in
the $N=2$ case the even dimension bosonic conserved charges exist only
for the $a=4$ hierarchy \cite{{Mat1},{Mat2}}. As pointed out in Ref. 14,
to learn whether the other
two $N=2$ hierachies admit an extension to $N=4$, perhaps under different
restrictions on the parameters $a^{ij}$, the construction of the
dimension 5 conserved charge for $N=4$ KdV would be crucial. In the
$N=2$ case it exists for all three super KdV hierarchies and is given
by different expressions in every case \cite{Mat2}. It is very
complicated
to construct such a quantity directly in HSS.
At the same time, for $N=2$ superfield computations there exist powerful
computer methods based on the package ``Mathematica'' \cite{comput}.
Keeping this in mind, it is tempting
to reformulate $N=4$ super KdV in terms of $N=2$ superfields.

This is one of the main purposes of the present paper. We rewrite the
$N=4$ super KdV in
$N=2$ superspace as a coupled system of equations for
a general dimension 1 superfield (this is just the $N=2$ KdV superfield)
and dimension 1 chiral and antichiral conjugated superfields.
This system involves two independent parameters which are the
components of the $SU(2)$ breaking tensor $c^{ijkl}$ in a fixed
$SU(2)$ frame. We explicitly construct the dimension $5$ and
$6$ conserved charges for this system (beside reproducing in the $N=2$
formalism the charges found in Ref. 14). They exist
if and only if the restrictions (\ref{consa1}), \p{consa} hold.
This is a very strong indication that $N=4$ KdV, with conditions
(\ref{consa1}), \p{consa}, gives rise to an integrable hierarchy and that
the latter is unique. One more argument in favour of the integrability
is that under the same restrictions on the parameters the $N=4$ super KdV
system is bi-hamiltonian. In this article we check this property also
for the
evolution equations associated with other conserved charges.
One more new result of this article  is the observation
that two inequivalent reductions of the same $N=4$ KdV to the
$N=2$ one are possible. They depend on how the $U(1)$
symmetry of the latter is embedded into the original $SU(2)$ group.
One of these reductions was described in Ref. 14 and it leads to
the $a=4$, $N=2$ KdV. The second one yields the $a=-2$, $N=2$ KdV.
Thus these two different $N=2$ KdV hierarchies prove to originate from
the single higher symmetry $N=4$ KdV hierarchy.

The paper is organized as follows. In Sec. II we recall, with
some further comments, the basic
points of our construction of $N=4$ super KdV in $N=4$, $1D$ HSS.
In Sec. III we rewrite $N=4$ KdV in ordinary $N=4$, $1D$ superspace
and then in $N=2$ superspace, and show the possibility of two
different reductions to $N=2$ super KdV. In Sec. IV the
dimension 4,5 and 6
conserved charges are constructed and shown to exist only with
the restrictions (\ref{consa1}), \p{consa}. Concluding remarks
are collected in Sec. V. Two Appendices contain some technical
details.

\setcounter{equation}{0}

\section{N=4 KdV in 1D harmonic superspace}
Here we recapitulate the salient features of $N=4$ super KdV equation
in the harmonic superspace formulation basically following Ref. 14.
We use a slightly different notation and add some comments.

\vspace{0.4cm}

\noindent {\bf 2.1. N=4 SU(2) SCA.} We started in Ref. 14 with
the $N=4$ $SU(2)$ superconformal algebra. In ordinary $N=4$, $1D$
superspace with  coordinates
\be \label{odSS}
Z^M \equiv
(x, \theta^i,\bar\theta_j)\;, (i,j=1,2)
\ee
this SCA is represented by the dimension 1
supercurrent $V^{ij}(Z) = V^{ji}(Z)$, $(V^{ij})^\dagger =
\epsilon_{ik}\epsilon_{jl} V^{kl}$, satisfying the constraints
(see, e.g., Ref. 19):
\begin{equation}
D^{(i}V^{jk)}=0\;,\ \ \bar D^{(i}V^{jk)}=0\;.
\label{ct1}
\end{equation}
Here
\be
D_i = \frac{\partial}{\partial \theta^i} - \frac{i}{2} \bar{\theta}_i
\frac{\partial}{\partial x}\;, \; \bar{D}^i = - \frac{\partial}{\partial
\bar{\theta}_i}
+ \frac{i}{2} \theta^i \frac{\partial}{\partial x} \;, \;\;
\{ D_i, \bar{D}^j \} = i\;\delta^i_j \partial \;, \; \{ D_i, D_j \}
=0\;, \label{ct1a}
\ee
the $SU(2)$ indices $i,j$ are raised and lowered by the antisymmetric
tensors $\epsilon^{ij}\;, \epsilon_{ij}$
($\epsilon^{ij}\epsilon_{jk} = \delta^i_k\;,
\epsilon_{12} = -\epsilon^{12} = 1$) and $(i_1...i_n)$ means
symmetrization (with the factor $1/n!$). It is straightforward
to check that
the constraints \p{ct1} leave in $V^{ij}$ only the following independent
superfield projections
\be
V^{ij}\;, \;\; \xi^{k} = D^i V^k_i\;,\;\; \bar \xi^{k} =-\bar D^i V^k_i
\;,\;\;T = \bar{D}^iD^k V_{ik}\;.
\label{ct1b}
\ee
The $\theta$ independent parts of these projections,
$w^{ij} (x),\;\xi^{l}(x),\; \bar \xi^{l}(x),\; T(x)$, up to inessential
rescalings coincide with the currents of $N=4\;SU(2)$ SCA:
the $SU(2)$ triplet of spin 1 currents generating
$SU(2)$ affine Kac-Moody subalgebra, a complex doublet of
spin 3/2 currents and the spin 2 conformal stress-tensor, respectively.
Superfield Poisson brackets between the $N=4$ $SU(2)$ supercurrents
leading
to the classical $N=4$ $SU(2)$ SCA for these component currents will be
presented below.

The same $N=4$ $SU(2)$ supercurrent admits an elegant reformulation
in the $N=4$, $1D$ harmonic superspace.

The latter is defined as
an extension of $\{ Z^M \}$ by the harmonic variables
$u^\pm_i$ describing a 2-sphere $\sim SU(2)/U(1)$
$$
\{ Z^M \} \Rightarrow \{ Z^M\;, \;u^{+i}\;, \;u^{-j} \}\;,
$$
\be
u^{+i}u_{i}^{-} = 1\;, \;\;\; u^{+}_iu^-_j - u^-_iu^+_j = \epsilon_{ij}
\label{ct1c}
\ee
(see Refs. 20 and 21 for details of the harmonic superspace approach).

In what follows we will need
the derivatives in harmonic variables which are given by
\bea D^{++} &\equiv & \pl^{++} = u^{+i}{\pl\over\pl u^{-i}}\;,\;
D^{--} \;\equiv \;\pl^{--} =u^{-i}{\pl\over\pl u^{+i}} \nonumber \\
D^0 &=& [ D^{++}, D^{--}] = u^{+i}{\pl\over\pl u^{+i}} -
u^{-i}{\pl\over\pl u^{-i}}\;. \label{ct1d} \eea
The operator $D^{0}$ measures the $U(1)$ charge of functions  on the
harmonic superspace. This charge is defined as the difference
between the
numbers of the  $+$ and $-$ indices. The preservation of this
$U(1)$ charge is one of
the basic postulates of the harmonic superspace approach.
It expresses the fact that the harmonic variables belong to the
sphere $S^2$ (actually contain two independent parameters)
and the harmonic superfields are functions on this sphere
as well. Let us notice that this
$U(1)$ charge commutes with the automorphism $SU(2)$ group
which acts on the doublet indices $i, j$.

Also, instead of $D^i\;,\;\bar{D}^j$ we will use their projections
on $u^{\pm i}$
\be
D^\pm=D^iu^\pm_i\;, \;\;
\bar D^\pm=\bar D^iu^\pm_i\;. \nn \\
\ee
Nonvanishing (anti)commutators of these projections with themselves and
with the harmonic derivatives $D^{++}$, $D^{--}$ are
\bea
\{ D^- ,\bar D^+ \} &=& i \pl \;,\  \{ D^+ ,\bar D^- \}\;=\;-i\pl \;,
\label{ct1e} \\
\left[ D^{++}, D^- \right] &=& D^+\;, \;\;
\left[ D^{--}, D^{+} \right]\;=\;
D^- \;. \label{ct1f}
\eea

We define now the $N=4$, $1D$ harmonic superfield $V^{++}(Z,u)$
subjected to the constraints
\bea
D^+V^{++} &=& 0\;, \;\;\;\;\bar  D^+V^{++}\;=\;0 \label{ct2} \\
D^{++} V^{++} &=& 0\;. \label{ct3}
\eea
(their consistency stems from the fact that the differential operators
in \p{ct2}, \p{ct3} are mutually (anti)commuting).
The harmonic constraint \p{ct3} implies that $V^{++}$
is a homogeneous function of degree 2 in $u^{+i}$
\be
V^{++} (Z,u) = V^{ij}(Z)\; u^+_iu^+_j \;. \label{ct3a}
\ee
Then, in view of the arbitrariness of $u^{+i}, \;u^{+j}$, the constraints
\p{ct2}
imply for $V^{ij}$ the original constraints \p{ct1}. Thus the
superfield $V^{++}$ obeying \p{ct2}, \p{ct3} represents the $N=4\;SU(2)$
conformal supercurrent in the harmonic $1D\;N=4$ superspace
(see also Ref. 22).

The constraints \p{ct2} can be viewed as Grassmann analyticity conditions
covariantly eliminating in $V^{++}$ the dependence on half of the original
Grassmann coordinates, namely, on their $u^-$ projections $\theta^{-} =
\theta^{i}u^{-}_i,\;\bar{\theta}^- = \bar{\theta}^iu^-_i$. So $V^{++}$ is
an {\it analytic} harmonic superfield living on an analytic subspace
containing only the $u^+$ projections of $\theta^i\;,\;\bar{\theta}^j$
\be
\{\zeta^M \} = \{ z, \theta^+, \bar{\theta}^+, u^+, u^- \}\;,
\label{analss}
\ee
$$
z = x - {i\over 2} \left( \theta^+ \bar \theta^- + \theta^- \bar \theta^+
\right), \; \theta^{\pm} = \theta^i u^{\pm}_i,\; \bar \theta^{\pm} =
\theta^i u^{\pm}_i.
$$
This harmonic analytic superspace is closed under the
action of $N=4$, $1D$ supersymetry (and actually under the transformations
of the whole $N=4\; SU(2)$ SCA, see below). Thus, one may construct
additional
superinvariants as integrals over this superspace.
This opportunity will be exploited when constructing the $N=4$ super KdV
hamiltonian and higher order conserved quantities.

In the analytic basis $\{ z, \theta^{\pm}, \bar \theta^{\pm},
u^{\pm}_i \}$
the covariant spinor derivatives $D^+$, $\bar D^+$ are reduced to the
partial derivatives
$$
D^+ = - \frac{\partial}{\partial \theta^-},\;\;\;
\bar D^+ = - \frac{\partial}{\partial \bar \theta^-}\;,
$$
and the conditions \p{ct2} indeed become Grassmann Causchy - Riemann
conditions stating the independence of $V^{++}$ on $\theta^-$,
$\bar \theta^-$ in this basis
$$
V^{++} = V^{++} (\zeta)\;.
$$
Now the irreducible components $w^{ij} (x),\;\xi^{l}(x), \bar \xi^{l}(x),
\; T(x)$ naturally appear in the $\theta^+$, $\bar \theta^+$ expansion of
$V^{++}$ as the result of solving the harmonic constraint \p{ct3}.
The analyticity-preserving harmonic derivative $D^{++}$ in the
analytic basis, when acting on
analytic superfields, is given by the expression
$$
D^{++} = \pl^{++} - i\theta^+ \bar \theta^+ \partial_z\;,
$$
and using this expression in Eq. \p{ct3} yields
\be
V^{++} (\zeta) = w^{ij} u^+_iu^+_j  - {2\over 3} \theta^+ \xi^k  u^+_k
+ {2\over 3} \bar \theta^+ \bar \xi^k u^+_k + \theta^+\bar \theta^+
\left( i \pl w^{ik} u^+_iu^-_k + {1\over 3} T  \right)\;,
\ee
where the numerical coefficients are inserted for agreement with the
definition \p{ct1b}.

It is easy to implement the superconformal $N=4$ $SU(2)$ group
as a group of
transformations in analytic superspace \p{analss}. Actually, there
exist two different realizations of this group in the superspace
\p{analss} \cite{{DS},{IS}} which yield as their closure the
``large'' $N=4$ $SO(4) \times U(1)$ superconformal
group \cite{{belg},{IKL}}.
The realization for which just $V^{++}$ serves as the supercurrent
can be written in the following concise form \cite{IS}
\bea
\delta z &=& (\pl^{--} D^{++} -2 ) \lambda\;,\;
\delta \theta^{+} \;=\; i \frac{\pl}{\pl \bar \theta^{+}} D^{++}
\lambda \;, \;
\delta \bar \theta^{+} = - i \frac{\pl}{\pl \theta^{+}} D^{++} \lambda
\;, \nonumber \\
\delta u^{+}_i &=& (D^{++} \pl \lambda) \;u^{-}_i \;\equiv\;
(D^{++} \Lambda^{0})\; u^-_i\;, \; \delta u^{-}_i \;=\; 0
\label{scgI}
\eea
Here, the analytic function $\lambda (\zeta)$ satisfies
the harmonic
constraint
\be
(D^{++})^2 \lambda (\zeta) = 0
\ee
and collects all the parameters of $N=4$ $SU(2)$ superconformal
transformations
\be
\lambda (\zeta) = \lambda + \lambda^{(ij)} u^{+}_i u^{-}_j +
\theta^{+} \varepsilon^i
u^{-}_i + \bar \theta^{+} \bar \varepsilon^{i}  u^{-}_i
+ i \theta^{+} \bar \theta^{+} \partial \lambda^{(ij)}
u^{-}_i u^{-}_j \;,
\ee
$\lambda (z), \varepsilon^i (z), \bar \varepsilon^i (z),
\partial \lambda^{(ij)} (z)$ being, respectively, the parameters of
the conformal, supersymmetry and $SU(2)$ affine transformations.

This realization of the $N=4$ $SU(2)$ superconformal group is fully
determined by the requirement that the harmonic
derivative $D^{++}$ transforms as
\be
\delta D^{++} = - (D^{++} \Lambda^{0})\; D^0\;.
\label{transfD}
\ee
The transformation law of $V^{++}$ is almost uniquely fixed
from the preservation of the harmonic constraint \p{ct3}:
\be
\delta V^{++} \simeq V^{++'}(\zeta') - V^{++}(\zeta) =
2 \Lambda^0 \;V^{++}  - {k\over 2}\; D^{++} \partial \Lambda^0\;,
\label{transfV}
\ee
where $k$ is a free parameter (its meaning will become clear soon).

In what follows
we will never actually need to know the explicit coordinate structure of
the analytic superspace and how $V^{++}$ is expressed there. We will only
make use of the constraints \p{ct2}, \p{ct3} and of some important
consequences of them, e.g.
\be
(D^{--})^3 V^{++} = 0\;, \;\; D^-(D^{--})^2 V^{++} =
\bar D^{-}(D^{--})^2
V^{++}
=0\;, \label{ct3b}
\ee
and those quoted in Appendix A.

After we have represented the $N=4\;SU(2)$ supercurrent as a harmonic
superfield $V^{++}$, it remains to write the Poisson bracket between two
$V^{++}$'s which yields the $N=4\;SU(2)$ SCA Poisson brackets for the
component currents. Surprisingly, this superfield Poisson bracket is
almost uniquely determined by dimensionality and compatibility with
the constraints \p{ct2}, \p{ct3}. It reads
\bea  \left\{ V^{++}(1),
 V^{++}(2)\right\} &=& {\cal D}^{(++|++)} \Delta (1 - 2) \nn \\
{\cal D}^{(++|++)} &\equiv & (D^+_1)^2(D^+_2)^2 \left(
\left[ \left({u^+_1u^-_2
\over u^+_1u^+_2}\right) - {1\over 2}D^{--}_2\right]
 V^{++}(2)
- {k\over 4}\pl_{2}\right),
\label{poi}\eea
where $\Delta(1-2)=\delta(x_1-x_2)\;(\theta^1-\theta^2)^4$ is the
ordinary $1D \;N=4$ superspace delta function and
$$
(D^+)^2 \equiv D^+ \bar D^+\;.
$$
We refer to Refs. 21
for more details on harmonic distributions. Note that the harmonic
singularity in the r.h.s. of \p{poi} is fake: it is cancelled after
decomposing the harmonics $u^{\pm i}_2$ over $u^{\pm i}_1$ with making
use of the completeness relation \p{ct1c} and the general formula
\p{decs} from Appendix A.

Using the algebra of spinor
and harmonic derivatives and also the completeness condition \p{ct1c},
one can check that the r.h.s of \p{poi} is consistent with the
constraints \p{ct2}, \p{ct3} with respect to both sets of arguments and
antisymmetric under the interchange $1\Leftrightarrow 2$. Note that
we should require the preservation of the harmonic $U(1)$ charge
independently for the points $1$ and $2$ in order to guarantee
that both sets of harmonic variables $u^\pm_{1\;i}$ and $u^\pm_{2\;i}$
parametrize the corresponding internal spheres $S^2$.

To be convinced that \p{poi} gives rise to the correct Poisson brackets
for the component
currents, we deduce from (\ref{poi})
the Poisson brackets of $SU(2)$ affine Kac-Moody currents.
After simple algebraic manipulations we obtain for $w^a \equiv
\sigma _i^{a\;j}w_j^{\;\;i}$ the familiar relation:
\bea
\left\{ w^{a}(1), w^{b}(2)\right\}= \epsilon^{abc}w^{c}(2)\;\delta (1-2)
- {k\over 2} \;\delta^{ab}\;\pl_{2}\delta(1-2) \;.\eea
All other currents can also be checked to satisfy the structure
relations of $N=4\;SU(2)$ SCA. We see that the central charge $k$ in
\p{poi} is the level of the affine $su(2)$ subalgebra.

It is straightforward to rewrite the Poisson
structure \p{poi} in ordinary $N=4$, $1D$ superspace.
There it looks much more complicated:
it involves intricate combinations of $SU(2)$ indices, etc. We
will quote
it in the next Section as an intermediate step in the derivation
of the $N=2$ superfield form of this structure.

Finally, we point out that the Poisson structure \p{poi} allows us to
write the $N=4$ superconformal transformation law of the
supercurrent in the following basis-independent form
\be \label{sconftrV}
\delta^* V^{++} (\zeta')= 4i\; \int [d\zeta^{-2}] \lambda (\zeta)
\left\{ V^{++}(\zeta), V^{++}(\zeta') \right\} \Rightarrow
\ee
\bea
\delta^* V^{++} (\zeta') &=& 2 (\partial \lambda) \;V^{++}  +
(2 \lambda - D^{--}D^{++} \lambda)
\; \partial V^{++}  - (D^{++} \partial \lambda) D^{--} V^{++}
\nonumber \\
&& + i (D^- D^{++} \lambda)\; \bar D^- V^{++}  - i
(\bar D^- D^{++} \lambda)\; D^- V^{++} - {k\over 2} D^{++}
\pl^2 \lambda \;,
\label{dettrV}
\eea
where $[d\zeta^{-2}] = dz [du] D^- \bar D^-$ is the measure of
integration over the analytic superspace (the integral over harmonics
is defined in the standard way: $\int [du] 1 = 1$ and the
integral of any symmetrized product of harmonics is vanishing \cite{a10}).
It is easy to see that this variation obeys the defining constraints
\p{ct2}, \p{ct3}. In the analytic basis of the harmonic superspace, it
becomes the active form of the variation \p{transfV}. The coefficient
before the inhomogeneous term in \p{transfV} has been chosen
for consistency with the fundamental Poisson structure \p{poi}. Note
that in deriving (2.24) from (2.23) and (2.21) we essentially
exploited the identity (A.6) from Appendix A.

It is interesting to note that the Poisson bracket \p{poi} can be used
to introduce the notion of primarity for analytic harmonic $N=4$
superfields.
Namely, let us consider a generalization of $V^{++}$, the
analytic superfields $L^{+l}$ subjected to the same harmonic constraint
\p{ct2}
$$
D^{++} L^{+l} = 0
$$
(they can be chosen real for $l= 2n$). The homogeneous $N=4$ $SU(2)$
superconformal transformation law of $L^{+l}$ unambiguously follows from
the preservation of this constraint
$$
\delta L^{+l} = l \Lambda^0 \;L^{+l}\;.
$$
This law can be equivalently reproduced by a formula of the type
\p{sconftrV}, with the following Poisson bracket between $V^{++}$ and
$L^{+l}$
\be \label{poiL}
\left\{V^{++}(1),  L^{+l}(2)\right\} =
{1\over 2} (D^+_1)^2(D^+_2)^2 \left(
\left[ l \left( { u^+_1u^-_2
\over u^+_1u^+_2} \right) - D^{--}_2 \right]
 L^{+l}(2)\;\Delta (1 - 2) \right)\;.
\ee
This bracket can be viewed as the manifestly supersymmetric definition
of $N=4$ $SU(2)$
primarity for the constrained analytic superfields $L^{+l}$ (at the
classical level). It would be of
interest to know whether one can define appropriate Poisson brackets
between the superfields $L^{+l}$ so that they form, together with \p{poi}
and \p{poiL}, a closed algebra providing an extension
(perhaps, nonlinear) of $N=4$ $SU(2)$ SCA.

\vspace{0.4cm}
\noindent{\bf 2.2. N=4 super KdV}. To deduce the
super KdV equation with the second
hamiltonian structure given by the $N=4\;SU(2)$ SCA in the form \p{poi}
we need to construct the relevant hamiltonian of the dimension 3.
The only requirement we impose {\it a priori} is that of $N=4$, $1D$
supersymmetry. The most general dimension 3
$N=4$ supersymmetric hamiltonian $H_3$ one may construct out
of $V^{++}$ consists of two pieces
\be
H_3 =\int [dZ]\;V^{++}(D^{--})^2V^{++}-i\int [d\zeta^{-2}]\;
c^{-4}(u)\;(V^{++})^3
\;.
\label{h3}
\ee
Here $[dZ]=dx[du]\;D^-\bar D^- D^+\bar D^+$ is the integration measure of
the full harmonic superspace. We see that
the $U(1)$ invariance of the integral over analytic subspace requires
the inclusion of the harmonic monomial
$c^{-4}(u)=c^{ijkl}u^-_iu^-_ju^-_ku^-_l$ which explicitly breaks $SU(2)$
symmetry. The coefficients $c^{ijkl}$  belong to the dimension 5 spinor
representation of $SU(2)$, i.e. form a symmetric traceless rank 2 tensor,
and completely break the
$SU(2)$ symmetry, unless $c^{-4}$ is of the
special form
\be  \label{strc4}
c^{-4}(u)=(a^{-2}(u))^2\;, \;\;\;
a^{-2}(u)=a^{ij}u^-_iu^-_j\;.
\ee
After taking off the harmonics this condition becomes
Eq. \p{consa1}. In this case, the symmetry
breaking parameter belongs to the dimension 3 (vector) representation of
$SU(2)$, and thus has
$U(1)$ as a little group. We point out that the presence of the trilinear
term in the hamiltonian is unavoidable if one hopes to eventually obtain
an integrable super KdV equation (it should be reduced in some limit to
the $N=2$ super KdV family which is integrable only providing the relevant
hamiltonian contains a trilinear term). Thus, one necessary condition
for the integrability of $N=4$ super KdV is that
$SU(2)$ is broken, at least down to its $U(1)$ subgroup.

Using the hamiltonian \p{h3}, we construct the relevant
evolution equation:
\be  V^{++}_t = \left\{ H, V^{++} \right\}.\ee
After some rather tedious but straightforward computations, it may
be cast into the following form:
\bea
 V^{++}_t&=&i \left( D^+ \right)^2 \left\{ {k\over 2}D^{--}V^{++}_{xx}
-\left[V^{++}(D^{--})^2V^{++}-{1\over 2}(D^{--}V^{++})^2\right]_x\right.\cr
&& \left. -{3\over 20}k A^{-4}(V^{++})^2_x+{1\over 2}A^{-6}(V^{++})^3
\right\}\;. \label{kdv4}
\eea
Here $A^{-4}$ and $A^{-6}$ are differential operators on the 2-sphere
$\sim SU(2)/U(1)$
\bea
A^{-4}&=&\sum_{N=1}^4(-1)^{N+1}c^{2N-4}{1\over N!}(D^{--})^N,\cr
A^{-6}&=&{1\over 5}\sum_{N=0}^4(-1)^{N}c^{2N-4}{(5-N)
\over (N+1)!}(D^{--})^{N +1}\;.\eea
We have used the notation:
\be c^{2N-4}={(4-N)!\over 4!}(D^{++})^Nc^{-4}, \ N=0\cdots 4\;.\ee

Equation \p{kdv4} is the $N=4$ $SU(2)$ super KdV equation we sought for.
It is easy
to check that its r.h.s satisfies the same constraints
\p{ct2}, \p{ct3} as the l.h.s. One might bring \p{kdv4} into a
more explicit form using the algebra
\p{ct1e}, \p{ct1f} (the first term takes then the familiar form
$-{k\over 2}\;V^{++}_{xxx}$), but for technical reasons it is convenient
to keep
the analytic subspace projector $\left( D^+ \right)^2$ in front of
the curly brackets in \p{kdv4}. The hamiltonian
\p{h3} and Eq.\p{kdv4} can be rewritten in ordinary
$N=4$ superspace (Sec. III), but
they look there very intricate, like the Poisson bracket \p{poi}. For
instance, the second term in \p{h3} would involve explicit $\theta$s, so
that it would be uneasy to see that it is supersymmetric. Thus, harmonic
superspace seems to provide the most appropriate framework
for a manifestly $N=4$ supersymmetric formulation of
$N=4$ super KdV equation. The last comment concerns the presence of the
$N=4\;SU(2)$ SCA central charge $k$ in \p{kdv4}. Making in \p{kdv4}
the rescalings
$t \rightarrow bt,\;V^{++} \rightarrow b^{-1}V^{++},\; c
\rightarrow bc$, we
can in principle change this parameter to any non-zero value. However,
in order to have a clear contact with the original
$N=4\;SU(2)$ Poisson structure \p{poi}, for the time being we prefer
to leave $N=4$ super KdV in its original form.

\vspace{0.4cm}
\noindent{\bf 2.3. Conserved charges.} As was mentioned in Introduction,
the $N=2$ super KdV equation is integrable
only for $a=4,\;-2,\;1$. Since the $SU(2)$ breaking tensor
$c^{ijkl}$ is a direct analog of the $N=2$ KdV parameter $a$
(and is reduced to it upon the reduction $N=4 \rightarrow N=2$,
see Sec. III), one may expect that
the $N=4$ super KdV equation is integrable only when certain
restrictions are imposed on this tensor. To see which kind of
restrictions arises, in \cite{DI} we required the existence of
non-trivial conserved charges for \p{kdv4} which are in involution
with the hamiltonian \p{h3}. Here we recall the results of that
analysis.

Conservation of the dimension 1 charge :
\be H_1=\int [d\zeta^{-2}]\; V^{++} \label{h1} \ee
imposes no condition on the parameters of the hamiltonian.

A charge with dimension 2 exists only provided the condition \p{strc4}
(\p{consa1}) holds. It reads:
\be H_2= i\int [d\zeta^{-2}]\; a^{-2}\;(V^{++})^2 \;.\ee
The conservation of this charge implies a stringent constraint
on $a^{ij}$, namely
\be s\equiv a^{+2}a^{-2}-(a^0)^2={1\over 2}\;a^{ij}a_{ij}=
-{10\over k}\;,
\label{ct4} \ee
where
$$
a^{+2} = D^{++} a^0 = {1\over 2}\;(D^{++})^2 a^{-2} = a^{ij}u^+_iu^+_j\;.
$$
This is just the second condition \p{consa} quoted in Introduction.
Note that with the convention \p{strc4} this condition implies for
$a^{ik}$ the following reality properties
\be
(a^{ik})^\dagger = - \epsilon_{ij} \epsilon_{kl} \;a^{kl}
\Leftrightarrow (a^{12})^\dagger = a^{12}, \;\; (a^{11})^\dagger =
- a^{22} \;.
\label{reality}
\ee
Assuming that the central charge $k$ is integer (if
we restrict ourselves to unitary representations of the $SU(2)$
Kac-Moody
algebra \cite{wit}), Eq. \p{ct4} means that $a^{ij}$ parametrizes some
sphere $S^{2} \sim SU(2)/U(1)$, such that the reciprocal of its radius
is {\it quantized}. It is interesting to explicitly find
the evolution equation produced by $H_2$ through the
hamiltonian structure \p{poi}
\be
V^{++}_{t'} = 3\left\{ H_2, V^{++} \right\} \Rightarrow
\label{ev2} \ee
\bea
V^{++}_{t'} &=& {i\over 2}\;(D^+)^2 \left\{ k\;\tilde{A}^{-2}
\;V^{++}_x -
3\;\tilde{A}^{-4} \left(V^{++} \right)^2 \right\}\;, \label{kdv2} \\
\tilde{A}^{-2} &=& a^0 \;D^{--} -{1\over 2}\;a^{+2} \;(D^{--})^2 \;,
\nonumber \\
\tilde{A}^{-4} &=& a^{-2}\; D^{--} -{1\over 3} \;a^0 \;(D^{--})^2
+ {1\over 18}\;
a^{+2}\; (D^{--})^3\;. \nonumber
\eea
(the factor $3$ in  \p{ev2} was chosen for further convenience).
This equation is the first non-trivial one in the conjectured
$N=4$ KdV hierarchy. As was recently noticed \cite{KS},
the $N=2$ counterpart of this equation can be interpreted as
a ``disguised'' form of the $N=2$ supersymmetric
extension of the nonlinear Schr\"odinger equation (NLS).
Thus, it is natural
to expect that Eq. \p{kdv2} is related in an anlogous way to the
$N=4$ extended NLS.

The last conserved charge we constructed in Ref. 14 is a
dimension 4 one $H_4$ (the dimension 3 conserved charge is the
$N=4$ KdV hamiltonian itself).
$H_4$ exists under the same restrictions \p{strc4},  \p{ct4}
(or, equally, \p{consa1}, \p{consa}) on  $c^{ijkl}$
and reads:
\be H_4=\int [dZ]\;a^{-2} V^{++}(D^{--}V^{++})^2
+{i\over 6}\int [d\zeta^{-2}] \left[{7\over 6}\;(a^{-2})^3 (V^{++})^4 -k
a^{-2} ( V^{++}_x)^2\right]. \label{ct5}
\ee
It is curious that it yields the same $N=4$ KdV equation \p{kdv4} via
the first hamiltonian structure associated with the Poisson bracket
\be
\left\{V^{++}(1), V^{++}(2)\right\}_{(1)}= i\beta
\left( a^{0}(1) - a^{+2}(1){u^-_1u^+_2\over u^+_1u^+_2}
\right)
(D^+_1)^2 (D^+_2)^2
\Delta(1-2) \label{poi1}\;.
\ee
Here, $\beta$ is an arbitrary real constant. This bracket is related to
the original one \p{poi} by the shift
\be V^{++}\longrightarrow V^{++}+ i\beta a^{+2}(u)\;.
\ee
Taking as a new hamiltonian
\be
H_{(1)}= -i {9k\over 4\beta}\;H_4\;,\label{ha1}
\ee
we reproduce \p{kdv4} as the hamiltonian flow:
\be V^{++}_t=\{H_{(1)},V^{++}\}_{(1)} \;. \label{kdv41} \ee
This comes about in a very non-trivial way,
since both the new Poisson bracket (\ref{poi1})
and the new hamiltonian (\ref{ha1}) are proportional to the
$SU(2)$ breaking parameter $a^{ij}$, while the super
KdV equation (\ref{kdv4}) includes terms containing
no dependence on $a^{ij}$.
The key point is that these terms appear in (\ref{kdv41})
multiplied by the factor
$$
-{k\over 10}\;s= -{k\over 20} \;a^{ij}a_{ij},
$$
which is independent of harmonic coordinates $u^\pm$ and is
constrained to be $1$ from the condition \p{ct4}.

Thus the conditions \p{strc4} and \p{ct4} (\p{consa1}, \p{consa})
are necessary not only
for the existence of the first non-trivial conservation laws for
Eq. \p{kdv4}, but also for it to be bi-hamiltonian. This property
persists for the evolution equations associated with other
conserved charges. For instance, with respect to the structure \p{poi1}
Eq. \p{kdv2} with $a^{ik}$ constrained by \p{consa}
has $H_3$ as the hamiltonian.

The presence of the bi-hamiltonian structure and the existence of
non-trivial
conserved charges are indications that $N=4$ KdV equation \p{kdv4}
with the restrictions \p{strc4}, \p{ct4} (\p{consa1}, \p{consa}) is
integrable, i.e. gives rise to a whole $N=4$ super KdV
hierarchy. Clearly, in order
to prove this, one should, before all, either find the relevant
Lax pair or
prove the existence of an infinite number of conserved
charges of the type
given above (e.g., by employing recursion relations implied by
the bi-hamiltonian property \cite{{Mag},{Opop}}). Unfortunately,
at present it is a very non-trivial and technically complicated
problem to
analyse these issues in full generality in the framework of harmonic
superspace. Even the direct construction of the next, dimension 5
charge $H_5$, turned out to be too intricate. In Sec. III we will
reformulate
$N=4$ KdV in $N=2$ superspace
where powerful computer methods for such calculations have been
developed. One thing which can be proven in a relatively simple
way in the framework of the HSS formalism is that
Eq. \p{strc4} (Eq. \p{consa1}) is a necessary condition for the
existence
of higher-order conserved charges for \p{kdv4}. We end the
present Section with the proof.

First of all, it is clear that after the reduction to $N=2$ such
charges should become those of the integrable $N=2$ super KdV equations
(see Sec. III for details of this reduction).
Any such charge of dimension, say, $l$ is known to contain in the
integrand a term $\sim V^l$, where $V$ is
the $N=2$ super KdV superfield ($N=2$ superconformal
stress-tensor) \cite{{Mat1},{Mat2}}.
These terms can only be obtained by reduction of analytic
integrals of the form
\be
\sim \int [d\zeta^{-2}]\; b^{-2(l-1)}\;  (V^{++})^l\;, \label{cc1}
\ee
where
\be
b^{-2(l-1)} = b^{i_1...i_{2(l-1)}} \;u^-_{i_1} ...
u^-_{i_{2(l-1)}}\;. \label{cc2}
\ee
If the corresponding charge is to be conserved, the highest
order contribution
to the time derivative of \p{cc1} (coming from the 3-d order term in the
r.h.s. of \p{kdv4}) should vanish separately. A simple analysis
shows that it is possible if and only if
\be
b^{-2(l-1)}
\sim (a^{-2})^{l-1}\;,\; \;c^{-4} = (a^{-2})^{2} \;. \label{cc3}
\ee
Note that in our previous paper \cite{DI} an erroneous statement
that this condition is necessary only for $l = 2n$ was made.

\setcounter{equation}{0}
\section{N=4 KdV in N=2 superspace}

\noindent{\bf 3.1. N=4 KdV and N=4 SU(2) SCA in ordinary N=4 superspace}.
We first rewrite Eq. \p{kdv4} in ordinary $N=4$ superspace, where it is
expressed as an equation for the
superfield $V^{ij}(Z)$ constrained by Eqs. \p{ct1}.
A straightforward calculation, that makes use of the identities
given in Appendix A, yields
\bea
V^{ij}_t &=& \left\{ -{1\over 2}\;k \;V^{ij}_{xx}
-2 \;V^{(il}_x V^{j)}_l +
{2i\over 3}\; T V^{ij} - {4i\over 9}\; \xi^{(i}\;\bar \xi^{j)}
- {3\over 10} \left( c^{klf(i}\;V_{kl}V_f^{j)} \right)_x
\right. \nonumber \\
&& + \left. {3\over 10}\; k \;c^{klf(i}\;V_{kl\;x}V^{j)}_f
 - {i\over 10}\; k \;c^{ijkl}
\left( T V_{kl} + {4\over 3} \xi_{(k} \bar \xi_{l)} \right) \right\}_x
- {3\over 10} \;c^{klfg} \;V_{kl} V_{fg} V^{ij}_x  \nonumber \\
&& - {3\over 5}\; c^{klfg} \;
V_{kl} \left( V^i_f V^j_g \right)_x
 - {i\over 5} \;c^{klf(i}
\left( B^{j)}_k V_{lf} + V^{j)}_k B_{lf} \right)\;.  \label{kdv44}
\eea
Here,
\be
B^{ij} \equiv T V^{ij} + {8\over 3}\; \xi^{(i}\bar{\xi}^{j)}\;,
\label{defw}
\ee
the irreducible superfield projections $\xi^k, \;\bar \xi^l, \;T$
were defined in \p{ct1b} and the subscript ``$x$'' corresponds as before
to $x$ -derivative.
We have verified that both sides of Eq. \p{kdv44} respect the
constraints \p{ct1}.

It is also straightforward to rewrite the Poisson bracket \p{poi}
in ordinary $N=4$ superspace
\be  \label{st4}
\left\{ V^{ij}(Z_1), V^{kl}(Z_2) \right\} =
- {\cal D}^{(ij|kl)}\;\Delta (1-2)\;,
\ee
\bea
{\cal D}^{(ij|kl)} &=& {1\over 4}
\left\{ \left[i\;V^{ik} \;D^{jl} \;\partial+
{1\over 3} \;V ^{ik} \;\epsilon^{jl} \;D^4 +
{i\over 2} \;\partial V^{kl} \;D^{ij}
\right. \right. \nonumber \\
&& + \left. \left. {1\over 6} \left( \bar \xi^k \;\epsilon^{il}
D^2 \;\bar{D}^j +
\xi^k \;\epsilon^{il} \bar{D}^2 \;D^j
\right) \right. \right. \nonumber \\
&& + \left. \left. {i\over 4}\;k\;
\left( \epsilon^{jl}\; D^{ik} \;\partial^2 - {i\over 3}
\; \epsilon^{ik}\;\epsilon^{jl} \;D^4 \;\partial \right) \right] +
\left( k \leftrightarrow l \right) \right\}
+ \left( i \leftrightarrow j \right)\;. \label{str4}
\eea
Here
\be
D^{ij} \equiv D^{(i} \bar{D}^{j)}\;, \;\; D^2 \equiv D^i D_i\;, \;\;
\bar{D}^2 \equiv \bar{D}_i \bar{D}^i\;, \;\; D^4 \equiv D^{ij} D_{ij}\;,
\label{defDD} \ee
and the differential operator \p{str4} is evaluated at the point $Z_2$.

The $N=4$ KdV hamiltonian \p{h3} and the other conserved charges presented
in the previous Section can also be appropriately rewritten. However,
it is not very enlightening to do so, because, as was already said above,
only those pieces of these charges which live in the whole harmonic
superspace retain a manifestly supersymmetric form after passing to the
standard $N=4$ superspace (e.g., the integrand in the first term in \p{h3}
becomes $\sim V^{ij}V_{ij}$). The ordinary $N=4$ superspace form
of the analytic harmonic superspace pieces explicitly includes $\theta$s.
Below we will rewrite these conserved quantities via $N=2$ superfields,
so that both kinds of terms will be represented as integrals over the same
$N=2$ superspace without explicit $\theta$s in the integrands.

\vspace{0.4cm}
\noindent{\bf 3.2. From N=4 to N=2.}
To make a reduction to $N=2$ superspace, we split the
$N=4$, $1D$ superspace \p{odSS} as follows
\be \label{N2ss}
\{ Z^M \} = (x, \theta, \bar{\theta}) \otimes (\eta, \bar{\eta}) \equiv
\{ Z^\mu \} \otimes (\eta, \bar{\eta})
\ee
with
\be
\theta \equiv \theta^1\;, \;\; \bar{\theta} \equiv \bar{\theta}_1\;,
\;\; \eta \equiv \theta^2\;, \;\;\bar{\eta} \equiv \bar{\theta}_2\;.
\ee
We also split the set of covariant spinor derivatives into those
acting in $N=2$ SS $\{Z^\mu \}$ and those acting on the extra spinor
coordinates $\eta, \bar{\eta}$
\bea
D_1 &\equiv& D\;, \;\; \bar{D}^1 \;\equiv \;\bar{D}, \;\;
D_2 \;\equiv \;d \;, \;\;
\bar{D}^2 \;=\; \bar{d}\;, \nonumber \\
\{D, \bar{D} \} &=& i\partial\;, \;\;\; \{d, \bar{d} \} = i\partial\;,
\label{covdN2}
\eea
(all other anticommutators are vanishing). Then we put the constraints
\p{ct1} into the form
\bea
DV^{22} &=& 0 \;, \;\;  dV^{11} \;= \;0\;, \;\;
dV^{12} \;= \; {1\over 2} DV^{11}\;, \;\; dV^{22} \;= \; 2DV^{12}\;,
\label{constrD} \\
\bar{D} V^{11} &=& 0\;, \;\; \bar{d} V^{22} \;=\; 0\;, \; \;
\bar{d} V^{12}  \;= \; -{1\over 2} \bar{D} V^{22}\;, \;\;
\bar{d} V^{11} \;=\; - 2 \bar{D} V^{12}\;.
\label{constrbar}
\eea
The first equations in the sets \p{constrD} and \p{constrbar} are most
essential.
They tell us that the superfields $V^{11}$ and
$V^{22} = (V^{11})^\dagger$ are chiral and anti-chiral in the
$N=2$ superspace. The remainder of
constraints and their consequences
\be
d\bar{d} V^{12} = \bar{D}D V^{12}\;, \;\; d\bar{d} V^{11} =
iV^{11}_x\;, \;\;
\bar{d} d V^{22} = iV^{22}_x
\label{conseq}
\ee
serve to express all the coefficient $N=2$ superfields
in the $\eta, \bar{\eta}$ expansion of $V^{12}$, $V^{11}$ and
$V^{22}$ in terms of spinor and ordinary derivatives of the lowest order
$N=2$ superfields
\bea
V(Z^\mu) &\equiv& V^{12}(Z^M)|_{\eta =0}\;, \; \Phi(Z^\mu) \;\equiv \;
V^{11}(Z^M)|_{\eta =0}\;, \; \bar{\Phi} (Z^\mu) \;\equiv \;
V^{22}(Z^M)|_{\eta=0} \label{defN2} \\
\bar{D} \Phi &=& 0\;, \;\; D \bar{\Phi} \;=\; 0\;. \label{chir}
\eea
Note that $V(Z^\mu)$ is not constrained by \p{constrD}, \p{constrbar}.

Thus, by going to $N=2$ superspace we have explicitly solved the
constraints \p{ct1} in terms of an unconstrained $N=2$ superfield $V$
and a pair of conjugate
chiral and anti-chiral  $N=2$ superfields $\Phi$, $\bar{\Phi}$.
Now it is clear how to obtain the $N=2$ superfield form of Eq. \p{kdv44}.
Its r.h.s obeys the same constraints \p{ct1} as the l.h.s,
so one should express all the $N=4$ spinor derivatives in the former
through $N=2$ spinor derivatives, by using the constraints in the
form \p{constrD}, \p{constrbar}. Then one puts $\eta = \bar{\eta} =0$ in
both sides of the equations obtained. Some useful relations are
\bea
\xi^1 &=& -{3\over 2} \;DV^{11}\;, \; \xi^2 \;=\; -3 \;DV^{12}\;, \;
\bar{\xi}^1 \;=\; -3\;
\bar{D} V^{12}\;, \nonumber \\
\bar{\xi}^2 &=& -{3\over 2}\; \bar{D} V^{22}\;, \; \;
T =\; 3 \;[ \;D, \bar{D} \;]\; V^{12}\;.
\label{reluse}
\eea

The last step needed to put $N=4$ KdV in
a convenient $N=2$ superfield form consists of choosing an appropriate
frame with respect to the global $SU(2)$ that acts on both the doublet
indices in \p{kdv44} and the doublet indices of the $N=4$
superspace grassmann coordinates. It is
easy to show that this frame can always be chosen so that only two real
components in the $SU(2)$ breaking tensor $c^{iklj}$ are non-zero
\be
c^{1212} \equiv {5\over 6}\; a \;, \;\; c^{1111} = c^{2222} \equiv
{5\over 6}\; b \;, \;\;
c^{1112} = c^{2221} = 0
\label{frame}
\ee
(the numerical factors were introduced for further convenience).
Note that this is still true if $c^{ijkl}$ is bilinear in the constant
vector $a^{ik}$ in accord with Eq. \p{strc4}. In this important case
\be \label{strc44}
c^{ijkl} = {1\over 3}\;(a^{ij}a^{kl} + a^{ik}a^{jl} + a^{il}a^{jk})
\;.\ee
For three independent $SU(2)$ fixations of $a^{ik}$
\bea
(a) \; a^{11} &=& a^{22} = 0\;, \;\; a^{12} \neq 0; \;\;\;\;\;\;
(b) \; a^{12} = 0\;, \;\; a^{11} = a^{22} \neq 0; \nonumber \\
(c)\; a^{12} &=& 0\;, \;\;\;a^{11} = -a^{22} \neq 0\;,
\label{fixa}
\eea
the components of $c^{ijkl}$ satisfy \p{frame}.
The values of $a$ and $b$ are given by
\bea
(a)\;a &=& {4\over 5}\; a^{12} a^{12}\;, \;\;b = 0; \;\;\;\;\;
(b)\; a = {2\over 5}\; a^{11}a^{11}\;, \;\;\; b = 3 \;a \nonumber \\
(c)\; a &=& -{2\over 5}\;a^{11}a^{22}\;, \;\; \; b = -3 \;a \;.
\label{fixac}
\eea
As we will see below, only these choices of the $SU(2)$ frame allow an
unambiguous reduction to $N=2$ KdV.

The whole tensor $c^{ijkl}$ (or $a^{ik}$ in the case \p{strc4},
\p{strc44})
can be restored by an appropriate $SU(2)$ rotation.
It is instructive
to describe how these $SU(2)$ rotations, which are manifest in the
original $N=4$
superspace, are realized on $N=2$ superfields $V$, $\Phi$,
$\bar{\Phi}$
\bea
\delta^* V &=& -\lambda_0 \left(
\theta \frac{\partial}{\partial \theta} -
\bar{\theta} \frac{\partial}{\partial \bar{\theta}} \right)
V
+\lambda_+ \left[ \Phi + D \;( \theta \Phi ) \right]
- \lambda_- \left[ \bar{\Phi} - \bar{D}\; ( \;\bar{\theta}
\bar{\Phi} \;) \right] \;, \nonumber \\
\delta^* \Phi &=& \lambda_0 \left(2 -
\theta \frac{\partial}{\partial \theta} +
\bar{\theta} \frac{\partial}{\partial \bar{\theta}} \right)
 \Phi  + \lambda_- \bar{D} \;(\; \bar{\theta} V \;) \;,
\nonumber \\
\delta^* \bar{\Phi} &=& \lambda_0 \left(-2 -
\theta \frac{\partial}{\partial \theta} +
\bar{\theta} \frac{\partial}{\partial \bar{\theta}} \right)
 \bar{\Phi}   + \lambda_+ D \;(\; \theta V \;
)\;.
\label{su2trans}
\eea
For completeness, we also give the transformation properties of the
superfields under the second complex supersymmetry,
implicit in the $N=2$ superfield notation
\bea
\delta^* V &=& {1\over 2}\;\epsilon^2 D\Phi +
{1\over 2}\;\bar{\epsilon}_2
\bar{D}\bar{\Phi}\;, \nonumber \\
\delta^* \Phi &=& 2\;\bar{\epsilon}_2 \bar{D} V\;, \;\;\;
\delta^* \bar{\Phi} \;= \; 2\;\epsilon^2 D V \;.
\label{susy2} \eea

After these preparatory steps and choosing,
for convenience, $k=2$ henceforth, we deduce
the $N=4$ $SU(2)$ KdV equation in a $N=2$ superfield form
as the following system of coupled evolution equations
\bea
V_t &=& - V_{xxx} + 3i\; \left( \left[ D, \bar{D} \right] V
\;V \right)_x
- {i\over 2}\;(1-a) \left( \left[ D,\bar{D} \right] V^2 \right)_x
-3a\; V_x V^2 \nonumber \\
&& + {1\over 4} \;(a-4) \left( \Phi_x\bar{\Phi} -
\bar{\Phi}_x \Phi
\right)_x + {i\over 2}\;(a-1) \left( D\Phi \bar{D}
\bar{\Phi} \right)_x
-{3\over 2}\;a \left( V\Phi \bar{\Phi} \right)_x \nonumber \\
&& + {1\over 8}\;b
\left( \Phi^2 - \bar{\Phi}^2 \right)_{xx}  -
{3\over 4}\;b \left[ V \left( \Phi^2 + \bar{\Phi}^2 \right)
\right]_x + {i\over 2}\;b \left[ D,\bar{D} \right]\left[ V\left(
\Phi^2 - \bar{\Phi}^2\right) \right] \label{eqV} \\
\Phi_t &=& -\Phi_{xxx} -{5\over 4}\;b\;\Phi_x \Phi^2 -\bar{D} \left[
6i\; \left(DV\; \Phi \right)_x - i\;(a+2) \;D\left( V\Phi \right)_x
\right]
\nonumber \\
&& + \bar{D} D \left[ 3i\;a \left( V^2\Phi +
{1\over 4}\;\Phi^2 \bar{\Phi} \right)
+ i\;b \left( V\bar{\Phi} \right)_x +
i\;b \left( V^2\bar{\Phi} +{1\over 4}\;
\bar{\Phi}^2 \Phi \right) \right] \label{eqphi} \\
\bar{\Phi}_t &=& -\bar{\Phi}_{xxx} -{5\over 4}\;b\;
\bar{\Phi}_x \bar{\Phi}^2 + D \left[
6i\; \left( \bar{D} V\; \bar{\Phi} \right)_x -
i\;(a+2) \;\bar{D}\left( V \bar{\Phi} \right)_x \right]
\nonumber \\
&& + D \bar{D} \left[ 3i\;a \left( V^2\bar{\Phi}
+ {1\over 4}\;\bar{\Phi}^2 \Phi \right)
- i\;b \left( V \Phi \right)_x + i\;b \left( V^2 \Phi  +{1\over 4}\;
\Phi^2 \bar{\Phi} \right) \right] \label{eqphi2}\;.
\eea
We have explicitly checked that Eqs. \p{eqV} - \p{eqphi2} are covariant
under the second hidden supersymmetry \p{susy2}. It is also obvious from
the form of Eqs. \p{eqphi}, \p{eqphi2} that they are consistent with the
$N=2$ chirality properties of $\Phi$, $\bar{\Phi}$.

Proceeding in a similar way, one can rewrite the second and first
Poisson bracket structures
\p{poi} and \p{poi1} in terms of the
$N=2$ superfields
\be
\left\{ V^A(1), V^B(2) \right\} =  {\cal D}^{AB} \Delta^{(2)} (1-2)
\label{poi2}
\ee
\bea
{\cal D}^{11} &=& {1\over 4} \left( i V\partial +
i \partial V - \bar D V \;D
- DV \; \bar D  + {k\over 4}\; [D,\bar D] \partial \right) \nonumber \\
{\cal D}^{12} &=&
{1\over 4} \left( i \partial \Phi + 2 \Phi \;\bar D D -
D\Phi \;\bar D \right)\;,\;
{\cal D}^{13} =
{1\over 4} \left( i\partial \bar{\Phi} + 2 \bar{\Phi} \;D \bar D -
\bar D \bar{\Phi} \;D \right)\;, \nonumber \\
{\cal D}^{21} &=& {1\over 4} \left( 2 \Phi \;D\bar D + D\Phi \;\bar D
\right)\;, \;\; {\cal D}^{23} =
\left( V\; D\bar D + DV \;\bar D - {k\over 4}\; D\bar D \partial \right)
\nonumber \\
{\cal D}^{31} &=& {1\over 4} \left( 2\bar{\Phi} \;\bar D D + \bar D
\bar{\Phi}\; D
\right)\;, \; {\cal D}^{32} = \left( V \;\bar D D + \bar DV \;D +
{k\over 4} \;\bar D D \partial \right),  \nonumber \\
{\cal D}^{22} &=& {\cal D}^{33} = 0
\label{poi2det}
\eea
\be
\left\{ V^A(1), V^B(2) \right\}_{(1)} =  {\cal D}_{(1)}^{AB}
\Delta^{(2)} (1-2)
\label{poi12}
\ee
\bea
{\cal D}_{(1)}^{11} &=& - {1\over 4}\beta a^{12} \;\partial\;,
\;\;\; {\cal D}_{(1)}^{12} \;=\; {i\over 2} \beta a^{11}\;
\bar D D\;, \;\;\;
{\cal D}_{(1)}^{13} \;=\; {i\over 2} \beta a^{22}\;D \bar D\;, \nn \\
{\cal D}_{(1)}^{21} &=& {i\over 2} \beta a^{11}\;D \bar D\;, \;\;\;
{\cal D}_{(1)}^{22} \;=\; 0\;, \;\;\;\;\;\;\;
{\cal D}_{(1)}^{23} \;=\; i \beta a^{12}\;D \bar D\;, \nn \\
{\cal D}_{(1)}^{31} &=& {i\over 2} \beta a^{22}\;\bar D D\;, \;\;\;
{\cal D}_{(1)}^{32} \;=\; i \beta a^{12}\;\bar D D\;, \;\;\;
{\cal D}_{(1)}^{33} \;=\; 0  \label{str12}
\eea
In  these formulas we made use of the condensed notation
$$
V^A \equiv ( V, \Phi, \bar \Phi)
$$
and defined the $N=2$ superspace delta function by
$$
\Delta^{(2)} (1-2) \equiv
\left( d_2\bar d_2 \Delta (1-2) \right) |_{\eta = \bar \eta = 0}
\;.
$$
The differential operators ${\cal D}^{AB}$, ${\cal D}^{AB}_{(1)}$ are
evaluated at the second point of $N=2$ superspace.

Using the relation between the $N=4$ and $N=2$ superspace integration
measures
\be
[dZ] = \mu^{(2)}\; [du]\; d \bar{d}\;,\; \;[d\zeta^{-2}] = -
\mu^{(2)}\; [du] \;[\;(\theta\bar \theta d\bar d -1)\; u^-_1 u^-_2 +
\theta d\; u^-_1 u^-_1 + \bar{\theta} \bar d\; u^{-}_2 u^-_2 \;],
\label{relmeas}
\ee
with
\be
\mu^{(2)} \equiv dx \;d\theta \;d\bar{\theta} = dx \;D \;\bar{D}\;,
\ee
and the constraints \p{constrD}, \p{constrbar}, it is also easy
to get the $N=2$ superfield form of the $N=4$ KdV hamiltonian \p{h3}
\be  \label{h32}
H_3  = \int \mu^{(2)} \left\{ 8 \bar D V D V + 2i \Phi_x \bar \Phi
+ {i\over 3}\;a \left( V^3 + 6 V\Phi\bar \Phi \right) + i\;b
\left( V\Phi^2 + V\bar \Phi^2 \right) \right\} \;.
\ee
Note that each of the three parts of \p{h32},
viz. those with the coefficients $a$ and $b$
and the remainder, are
separately invariant with respect to
the hidden supersymmetry transformations \p{susy2}. At the same time,
only the first piece (containing no dependence on $a$ and $b$)
respects the invariance under the $SU(2)$ transformations \p{su2trans}
(it comes from the first integral in the original expression \p{h3}).
Applying \p{su2trans} to the terms proportional to $a$ and $b$, one can
restore all five components of the initial $SU(2)$ breaking
tensor $c^{iklj}$. These will appear with appropriate
$N=4$ super-invariant combinations of $V$, $\Phi$, $\bar \Phi$ and
derivatives of the latter.

It is a straightforward exercise to rederive Eqs. \p{eqV} -
\p{eqphi2} as the evolution equations with respect to the
$N=2$ superfield hamiltonian structure \p{poi2}, \p{poi2det}, \p{h32}
\be
V^A_t = {\cal D}^{BA} \frac{\delta H_3}{\delta V^B}\;.
\label{evkdv42}
\ee

As was mentioned in Sec. II, the same $N=4$ super KdV equation, provided
the constraints \p{consa1}, \p{consa} hold (their form in the
$N=2$ notation
will be discussed in the next Subsection), can be regarded as an
evolution
equation with respect to the first
Poisson structure \p{poi1}, \p{poi12}
with $H_4$ as the hamiltonian, Eq. \p{kdv41}. In $N=2$ superfield
language, this form of $N=4$ KdV is as follows
\be
V^A_t = -i {9\over  2 \beta}\;{\cal D}_{(1)}^{BA} \frac{\delta H_4}
{\delta V^B}\;.
\label{evkdv421}
\ee
The $N=2$ superfield form of the conserved charge $H_4$ \p{ct5}
will be given below (Sec. IV).

\vspace{0.4cm}

\noindent{\bf 2.4. Reduction to $N=2$ super KdV and the integrability
conditions.} The first line of Eq. \p{eqV}, up to unessential
redefinitions, is just the r.h.s. of
$N=2$ super KdV equation \cite{{Mat1},{Mat2}}, with the parameter
$a$ related to the $SU(2)$ breaking tensor of $N=4$ super KdV as
\be
a = {6\over 5}\;c^{1212}\;.
\ee
Thus, the reduction to $N=2$ KdV is obtained by putting
\be \label{condN2}
\Phi = \bar{\Phi} = 0
\ee
in Eqs. \p{eqV} - \p{eqphi2}. As a result of the reduction, one gets
\be \label{kdvn2}
V_t = - V_{xxx} + 3i\; \left( \left[ D, \bar{D} \right] V \;V \right)_x
- {i\over 2}\;(1-a) \left( \left[ D,\bar{D} \right] V^2 \right)_x
-3a\; V_x V^2 \;.
\ee
This equation is related to the standard form of $N=2$ KdV equation
given in Refs. 8 and 9 via the redefinitions
\bea
V &=& \tilde{V}\;, \;\partial_x = i \tilde{\partial}_x\;, \;
\partial_t  = -i \tilde{\partial}_t\;, \;
D  = {1\over 2} \left( D_1 + i D_2 \right)\;, \;\bar D =
-{1\over 2} \left( D_1 -iD_2 \right)\;, \nn \\
D_1^{\;\;2} &=& D_2^{\;\;2} \;=\; \tilde{\partial}_x\;. \label{redef}
\eea

We should point out that the above reduction is consistent because Eqs.
\p{eqphi} and \p{eqphi2} are homogeneous in $\Phi$, $\bar{\Phi}$ and,
for this reason, condition \p{condN2} together with Eq. \p{kdvn2}
yield a particular solution of
the original set \p{eqV} - \p{eqphi2}. The superfield $V$ satisfies
Eq. \p{kdvn2} and is unconstrained otherwise.
All the conserved charges of $N=4$ KdV become conserved charges of
$N=2$ KdV in the reduction limit.

This is not the case for any {\it other}
choice of the $SU(2)$ frame beside those leading to \p{frame}.
This is because for non-zero $c^{1112}$ and $c^{2221}$ there appear
extra pieces in the equations for $\Phi$, $\bar{\Phi}$, that
{\it do not vanish} after the reduction \p{condN2}. For example, for
Eq. \p{eqphi} these pieces are as follows
\be \label{extra}
\triangle \Phi_t = -{2\over 5}\;i\;c^{1112}\;\bar{D}D \left[
3 \;V V_x + 4 \;V^3 \right]\;.
\ee
In this case, the reduction to $N=2$ KdV is inconsistent, because the
superfield $V$ becomes constrained in the $N=2$ KdV limit (the r.h.s.
of Eq. \p{extra} should vanish after imposing \p{condN2}).
To avoid confusion, we mention that the systems associated with these
other choices of the $SU(2)$ frame are simply other ``$SU(2)$ gauges''
of the same $N=4$ super KdV equation.
They can be rotated into Eqs. \p{eqV} - \p{eqphi2} by an appropriate
$SU(2)$ transformation. Only in the $N=2$ KdV limit, where
$SU(2)$ covariance gets broken, different choices of the $SU(2)$ frame
turn out to lead to inequivalent systems.

Now let us see which values of $a$ correspond to the restrictions
\p{strc4}, \p{ct4} (or \p{consa1}, \p{consa}) that are required
for $N=4$ KdV to be integrable. According to the
reasonings just mentioned,
only three directions of the $SU(2)$ vector $a^{ij}$, summarized in
Eq. \p{fixa}, allow for an unambigous reduction to $N=2$
super KdV. Indeed, only under this choice the components
$$
c^{1112} = - (c^{2221})^\dagger = a^{11}a^{12}
$$
are zero. Then, substituting the relations \p{fixa} into \p{ct4} we find
three cases for which $N=4$ super KdV in the $N=2$ superfield form
\p{eqV} -
\p{eqphi2} is expected to be integrable
\be \label{valuesab}
(a)\; a= 4, \;\; b=0; \;\;\; \;(b)\; a = -2, \;\; b= -6; \;\;\; (c)\;
a= -2, \;\; b= 6\;.
\ee
In the full $N=4$ case these possibilities are all equivalent
since they are related by $SU(2)$ rotations. Nevertheless,
they yield inequivalent
systems upon the reduction \p{condN2}. Remarkably,
{\it these are precisely two integrable} $N=2$ {\it KdV hierarchies},
{\it the}
$a=4$ {\it and} $a=-2$ $N=2$ {\it KdVs } \cite{{Mat1},{Mat2}}.

Thus, the single $N=4$ $SU(2)$ super KdV equation \p{kdv4} (or its
equivalent
forms \p{kdv44} and \p{eqV} - \p{eqphi2}) with the restrictions
\p{strc4} and \p{ct4} (\p{consa1}, \p{consa}) embodies as
particular solutions two of the three
integrable inequivalent $N=2$ super KdV equations. Below we will
explicitly construct the $N=2$ superfield form of the
dimension 5 and 6 conserved charges
for $N=4$ KdV and show that they exist only for the values of the
parameters
$a$ and $b$ listed in Eq. \p{valuesab}. This is a strong evidence that
a unique $N=4$ $SU(2)$ KdV hierarchy exists, yielding by reduction
the $a=4$ and $a=-2$, $N=2$ KdV hierarchies \p{condN2}.
Reversing the argument, we conclude that only these two $N=2$ KdV
hierarchies can be promoted to the $N=4$ $SU(2)$ KdV hierarchy. It is
worth noting that in this respect the latter is
complementary to the $N=3$ super KdV one \cite{Yung} which yields,
upon the reduction to $N=2$ superspace, the $a=1$, $N=2$ KdV.

In the rest of this Section we discuss how to recover the restrictions
\p{consa}, \p{consa1} directly at the level of the $N=2$ superfield
formulation, starting from the
$N=2$ superfield system \p{eqV} - \p{eqphi2}, with the
parameters $a$ and $b$ restricted to the values \p{valuesab}
by some reasoning (e.g., coming
from the study of higher-order conserved quantities). The only extra
assumption will be that the parameters $a$ and $b$ correspond to a
$SU(2)$ fixed form of some constant tensor $c^{iklj}$
in accordance with the definition \p{frame}.
In other words, we assume that the system \p{eqV} - \p{eqphi2} still
``remembers'' about its manifestly $SU(2)$ covariant and
$N=4$ supersymmetric origin.

First of all, computing two independent invariants of $c^{iklj}$,
$$
{\cal A} \equiv c^{ijkl}c_{ijkl}, \;\; {\cal B} \equiv
c^{ik}_{\;\;\;jl} c^{jl}_{\;\;\;ft}
c^{ft}_{\;\;\;ik}\;,
$$
for three options in \p{valuesab}, we find that in all cases the
invariants take the same values
\be
{\cal A} = {2\over 3}\; 10^2\;, \;\;\;\;
{\cal B} = - {2\over 9}\; 10^3\;, \label{valuesAB}
\ee
 from which it follows that the above choices represent the same tensor
$c^{iklj}_0$ in different $SU(2)$ frames (up to possible discrete
reflection-type transformations of $c^{ikjl}$ ). Further,
according to the Lemma proved in Appendix, the necessary and sufficient
conditions for $c^{ijkl}$ to have the special form \p{consa1} are the
following two ones
\be
{\cal A}^3 = 6\;{\cal B}^2\;, \;\;\; {\cal B} < 0\;.
\label{condvec}
\ee
The values \p{valuesAB} satisfy these criterions, from which follows
the constraint \p{consa1} for $c^{iklj}_0$.
 From \p{consa1} and \p{frame} we find
\be
a = {2\over 5} \;\left( 2a^{12} a^{12}  + a^{11}a^{22} \right)\;, \;\;
b = {6\over 5}\; a^{11} a^{11} = {6\over 5}\; a^{22} a^{22}\;, \;\;
a^{12} a^{11} = a^{12} a^{22} = 0\;.
\ee
Then, for the three options in \p{valuesab}, we have the following
solutions for $a^{ik}$
\bea
(a)\; a^{12} &=& \pm \sqrt{5}, \;\;a^{11} \;=\;
a^{22} \;=\; 0; \;\;\; \;
(b)\; a^{12} \;=\; 0, \;\; a^{11} \;=\; a^{22} \;=\;
\pm i \sqrt{5}; \nn \\
(c)\;
a^{12} &=& 0, \;\; a^{11} \;=\; - a^{22} \;=\; \pm \sqrt{5}\;.
\label{valuesij}
\eea
In all these three cases
\be
|a|^2 = - a^{ij}a_{ij} = 2 \;(a^{12} a^{12} - a^{11} a^{22}) = 10
\label{consak2}
\ee
that is precisely the constraint \p{consa} at $k=2$. Note that
the reconstruction of the vector $a^{ik}$ from the known
$c^{ijkl}_0$
is unique modulo some reflections of $a^{ik}$, as is seen from the
explicit solution \p{valuesij}.

Finally, we note that, when analyzing the integrability
properties of $N=4$ super KdV in the $N=2$ superfield formulation, we
actually do not need to keep track of all these subtleties concerning
the relation between $c^{ijkl}$ and $a^{ik}$, etc. One can forget about
the $N=4$ superfield origin of the system \p{eqV} - \p{eqphi2} and
view it as some two-parameter extension of $N=2$ KdV equation. Then
the specific values \p{valuesab} of the parameters $a$ and $b$ come out
as the values at which this system possesses higher-order conserved
charges
and is bi-hamiltonian (see next Section). Of course, in order to see
that the three options in Eq. \p{valuesab} are actually
equivalent to each other, one should take into account the fact
that the system
\p{eqV} - \p{eqphi2} respects a hidden $SU(2)$ symmetry, or,
eqivalently,
admits a manifestly $N=4$ supersymmetric and $SU(2)$ covariant
description discussed in Sec. II. The above discussion was aimed
just at carefully clarifying the links between this latter description
and the $N=2$ superfield one.

\setcounter{equation}{0}
\section{Conserved charges in the N=2 superfield formulation}

In this Section we put into an $N=2$ superfield form all the $N=4$
super KdV conserved charges given in Ref. 14 and Sec. II and present
two new ones: $H_5$ and $H_6$. We find that all these charges exist
under the same restrictions \p{valuesab} which, as was discussed in the
end of previous Section, actually amount to the original constraints
\p{consa1}, \p{consa}.

\vspace{0.4cm}
\noindent{\bf 4.1 The charges $H_1$ and $H_2$.} In order to find
the $N=2$ superfield representation of the conserved charges initially
written as integrals over $N=4$ HSS and its analytic subspace, we
proceed in the same way that was used to get the $N=2$ superfield
form of $H_3$,
Eq. \p{h32}. Namely, we make use of the relations \p{relmeas} and
\p{constrD}, \p{constrbar} and do the harmonic integrals in the end.

The charge $H_1$ \p{h1} is of the same form as in the $N=2$ KdV case
\be
H_1 = -2 \;\int \mu^{(2)} V \;. \label{h12}
\ee

Starting with $H_2$, non-trivial contributions of the superfields
$\Phi$, $\bar \Phi$ come out
\be
H_2 = {4i\over 3} \;\int \mu^{(2)} \left\{ a^{12}
\left( V^2 + {1\over 2}
\Phi \bar \Phi \right) - a^{11} \;V\bar \Phi -
a^{22} \;V\Phi \right\} \;.
\label{h22}
\ee
Like in $H_3$, three terms in \p{h22} are separately invariant under
the hidden supersymmetry \p{susy2} but are mixed by the $SU(2)$
transformations \p{su2trans}. Assuming for the moment that the
coefficients
$a^{ik}$ are {\it arbitrary}, we have checked the conservation of
\p{h22}
with respect to Eqs. \p{eqV} - \p{eqphi2}, both ``by hand'' and
using the
computer, and found $(H_2)_t$ to vanish under the following conditions
\bea
(a)\; a &=& 4\;, \;b = 0\;, \;\; a^{12} \neq 0\;, \;
a^{11} = a^{22} = 0\;; \nn \\
(b)\; a &=& -2\;, \;b = -6\;, \;\;a^{12} = 0\;, \;
a^{11} = a^{22} \neq 0\;; \nn \\
(c)\;
a &=& -2\;, \;b = 6\;, \;\; a^{12} = 0\;, \;
a^{11} = - a^{22} \neq 0\;.
\label{valuesah2}
\eea
Keeping in mind the discussion in the end of previous Section
(see Eqs. (3.42)), these solutions, up to relative scaling factors,
precisely correspond to the conditions \p{consa1}, \p{consa}
found from the computations in HSS.

This is the appropriate place to give the $N=2$ superfield form of
the evolution equation \p{kdv2} associated with  the hamiltonian $H_2$.
It can be obtained either by a direct transition to $N=2$ superfields
in \p{kdv2} or using the $N=2$ Poisson structure \p{poi2}, \p{poi2det}
\be
V^A_{t'} = 3 \;{\cal D}^{BA}\; \frac{\delta H_2}{\delta V^B} \;.
\label{evkdv22}
\ee
Both these equivalent ways yield the same $N=2$ superfield system
\bea
V_{t'} &=& a^{12} \left( i[D,\bar D] V -2 V^2 - \Phi\bar \Phi \right)_x
+ a^{11}
\left( {1\over 2} \bar \Phi_{xx} + 2\;(V\bar \Phi)_x + i\;[D,\bar D]
(V\bar \Phi) \right) \nn \\
&& -
a^{22}
\left( {1\over 2} \Phi_{xx} - 2\;(V \Phi)_x + i\;[D,\bar D]
(V \Phi) \right) \nn \\
\Phi_{t'} &=& i \bar D D \left\{ a^{12} \left( \Phi_x + 4 \;
V\Phi \right)
- 2 a^{11} \left( V_x + V^2 +{1\over 2}\;\Phi\bar \Phi \right)
- {3\over 2}\;
a^{22}\; \Phi^2 \right\} \nn \\
\bar \Phi_{t'} &=& - i D \bar D \left\{ a^{12} \left( \bar \Phi_x -
4\; V \bar \Phi \right)
- 2 a^{22} \left( V_x - V^2  - {1\over 2}\;\Phi\bar \Phi \right) +
{3\over 2}\; a^{11}\; \bar \Phi^2 \right\}\;. \label{kdv22}
\eea
This system can be derived in one more way, via the first
Poisson structure \p{poi12}, \p{str12} with $H_3$ as the hamiltonian.
However, this is possible only under the constraints \p{consa1},
\p{consa}. Requiring the equations
\be
V^A_{t'} = {\cal D}^{BA}_{(1)}\; \frac{\delta H_3}{\delta V^B}
\label{evkdv21}
\ee
to coincide, up to an overall renormalization factor,
with Eqs. \p{kdv22}
immediately leads to the restrictions \p{valuesah2} and, hence, to
\p{consa1}, \p{consa}. Thus, like in the case of the $N=4$ KdV equation,
the system \p{kdv22} is bi-hamiltonian only provided the basic
conditions \p{consa1}, \p{consa} hold. Of course, these conditions
can be also deduced
by demanding $H_3$ to be conserved with respect to Eqs. \p{kdv22} with
for the moment arbitrary coefficients $a^{ik}$, viz.
\be
\{ H_2, H_3 \} = 0\;. \label{commh23}
\ee
Clearly, this is equivalent to demanding $H_2$ to be conserved with
respect to $N=4$ KdV Eqs. \p{eqV} - \p{eqphi2}.

Finally, we observe that in the $N=2$ KdV limit \p{condN2} $H_2$
is non-zero for the option $(a)$ in \p{valuesah2} and identically
vanishes in the two other cases. As we will see, this property
persists for the $N=4$ KdV
charges $H_4$ and $H_6$. It reflects the fact that the even dimension
bosonic conserved quantities exist only for the
$a=4$, $N=2$ KdV, but not for the $a=-2$ one \cite{Mat2}.

\vspace{0.4cm}
\noindent{\bf 4.2 The charges $H_4$, $H_5$ and $H_6$.} We have found the
$N=2$ superfield form of the conserved charge $H_4$ in two ways:
first starting from the harmonic
superspace expression \p{ct5} and, second, constructing the most
general
dimension 4 expression directly in $N=2$ superspace and then checking
under which restrictions on the coefficients it is conserved with
respect to
Eqs. \p{eqV} - \p{eqphi2} (in doing this, we made use of the computer).
Both ways lead to the same answer.

Proceeding in the first way and representing $H_4$ as
\be
H_4 = \int \mu^{(2)} \left( {\cal H}_4^{I} + {\cal H}_4^{II} +
{\cal H}_4^{III} \right),  \label{N2h4}
\ee
where the three pieces in the integrand precisely correspond to the
three terms in the expression \p{ct5}, we get
\bea
{\cal H}_4^{I}  &=& -{4\over 3} \left\{ a^{12} \left( 4 \;V DV
\bar D V
- V D \Phi \bar D \bar \Phi \right) + a^{11} \left( 2\;\bar \Phi
\bar D V D V + {i\over 4}\; \bar \Phi^2 \Phi_x \right)
\right. \nonumber \\
&& + \; \left. a^{22} \left( 2 \;\Phi \bar D V DV -
{i\over 4} \;\Phi^2 \bar
\Phi_x \right) \right\} \nonumber \\
{\cal H}_4^{II}  &=&  {i\over 36} \left\{ {2\over 5}
\left( 2\;(a^{12})^3 +
3 a^{12} a^{11} a^{22} \right) \left( 8\;V^4 +
3 \;\Phi^2 \bar \Phi^2
+ 24 \;V^2 \Phi \bar \Phi \right)
\right. \nn \\
&& + \; \left. 4\;(a^{11})^2 a^{12} \left( \bar \Phi^3 \Phi +
6\;V^2 \bar
\Phi^2 \right) + 4 \;(a^{22})^2 a^{12} \left( \Phi^3 \bar \Phi +
6 \;V^2 \Phi^2 \right) \right.
\nn \\
&& - \; \left. 8 \;(a^{11})^3 V \bar \Phi^3  -
{8\over 5} \left( 4 (a^{12})^2 a^{11} + (a^{11})^2 a^{22}
\right) \left( 4\;V^3 \bar \Phi +
3 \;\bar \Phi^2 \Phi V \right) \right.
\nn \\
&& - \; \left. 8\;(a^{22})^3 V \Phi^3 - {8\over 5}
\left( 4 (a^{12})^2 a^{22} + (a^{22})^2 a^{11}
\right) \left( 4\;V^3 \Phi + 3 \;\Phi^2 \bar \Phi V \right)
\right\} \nn \\
{\cal H}_4^{III} &=& {4i\over 9} \left\{ a^{11}\;V_x \bar \Phi_x
+ a^{22}\;V_x \Phi_x - a^{12}  \left( V_xV_x
+ {1\over 2} \;\Phi_x\bar \Phi_x \right) \right\} \;.
\label{N2h4det}
\eea

On the other hand, the results of the second calculation can be
summarized as follows
\begin{eqnarray}
\hat{H}_4 & = & \int \mu^{(2)} \left( a_1\;  V^4 +
                     ia_2\;  V^3 \Phi +
        ia_3 \; V^3 {\bar \Phi} +
        3 a_1 \; V^2 \Phi {\bar \Phi} + \frac{3a_1}{8} \;
        \Phi^2 {\bar \Phi}^2   \right.  \nonumber \\
& &   + \; \left. i \frac{5a_3}{4}\;  V \Phi^3 +
        i \frac{3a_2}{4}\;  V \Phi^2 {\bar \Phi} +
       i \frac{3a_3}{4}\;  V \Phi {\bar \Phi}^2 +
       i \frac{5a_2}{4}\;  V {\bar \Phi}^3  \right. \nonumber \\
& &    -\; \left. i \frac{3a_{1}}{2} \; V^2 [ D, {\bar D}] V +
        i \frac{3a_{2}}{2}\;  V_x V \Phi -
        \frac{3a_{2}}{2} \; [ D, {\bar D} ] V  V \Phi
\right. \nonumber \\
& &    - \; \left. i\frac{3a_{3}}{2}\;  V_x V {\bar \Phi} -
        \frac{3a_{3}}{2} \; [D, {\bar D}] V  V {\bar \Phi}
        - i \frac{3a_{1}}{2}\;  V  D \Phi {\bar D} {\bar \Phi}
       - i \frac{3a_{2}}{4} \; {\bar \Phi} \Phi \Phi_x
\right. \nonumber \\
& &   + \;\left. i \frac{3a_{3}}{4} \; \Phi {\bar \Phi}
{\bar \Phi}_x +
       \frac{a_1}{2}\;  V V_{xx}  - i \frac{a_2}{2}\;
       V \Phi_{xx} -
       i \frac{a_3}{2}\;  V {\bar \Phi}_{xx} +
       \frac{a_1}{4}\;  \Phi {\bar \Phi}_{xx} \right) \label{N2h2com}
\end{eqnarray}

\begin{center}
\begin{tabular}{|c|c|c|c|} \hline
           & $a_1$ & $a_2$  & $a_3$ \\ \hline
a=4, b=0   & 1     & 0      & 0   \\ \hline
a=-2, b=-6 & 0     & -1/2     & -1/2 \\ \hline
a=-2, b=6  & 0     & 1/2      & -1/2   \\ \hline
\end{tabular} \vspace{0.5cm} \\
{\bf Table 1}
\end{center}
where in Table 1 we listed the values of the coefficients and
parameters $a$ and $b$ for which the charge \p{N2h2com} is conserved.
We stress that there are only these three solutions.
Substituting into Eqs. \p{N2h4det} the values \p{valuesij} of $a^{ij}$
which correspond to three different choices of the parameters
$a$ and $b$
in Table 1, we find that the relevant
$H_4$ and $\hat{H}_4$ differ (modulo full derivatives) merely by
unessential scaling factors. These factors are not fixed
by requiring the conservation of the $N=4$ KdV charges in the
$N=2$ superfield formalism and can always be chosen so as to achieve
the full coincidence between $H_4$ and $\hat{H_4}$. Thus,
the independent $N=2$ superfield calculation entirely confirms the
conclusions about $H_4$ made in our previous paper \cite{DI}
in the framework of the HSS formalism.

Having at our disposal the explicit $N=2$ superfield form of $H_4$
we can check the first hamiltonian structure representation
\p{evkdv421} for $N=4$ KdV system  \p{eqV} - \p{eqphi2}. Like
in the case of the set
\p{kdv22}, the necessary conditions for the existence of such a
representation are the above constraints on
the parameters $a$ and $b$. Actually, an alternative
and technically more simple way to obtain \p{N2h2com} with the
coefficients from Table 1 is to start from the most general
$N=2$ superfield expression for $\hat{H}_4$ and to require it to
reproduce Eqs. \p{eqV} - \p{eqphi2} via the Poisson structure
\p{poi12}, \p{str12}.

Note that for the second and third lines in Table 1, the charge $H_4$
identically vanishes in the $N=2$ KdV limit \p{condN2} in
accordance with
the absence of the even dimension bosonic conserved charges for the
$a=-2$, $N=2$ KdV hierarchy.

Let us now present the conserved charge $H_5$. As was already mentioned,
it is a very complicated technical problem to construct it directly in
the harmonic superspace formalism. This becomes feasible in the $N=2$
superfield approach due to the possibility to use a computer.
We start from the most general dimension 5 $N=2$ superfield
expression for $H_5$ with undetermined coefficients and then
examined the restrictions imposed on these coefficients by the
conservation condition $(H_5)_t = 0$. Like in the case of the
lower-dimension charges, we have found only three solutions
\bea
H_5 &=& \int \mu^{(2)} \left\{\frac{i}{4} \Phi \bar{\Phi}_{xxx} -
V [D, \bar{D}] V_{xx} -
i a_1 V^2 V_{xx}
+ 2i V [D,\bar{D}] V [D, \bar{D}]V  \right.  \nonumber \\
&&  +\; \left.  ia_{2} V\Phi \Phi_{xx} + \frac{ia_{2}}{2} V
\Phi_{x} \Phi_{x} + ia_{3} V\Phi_{xx} \bar{\Phi}
+i a_{4} V \Phi_x \bar{\Phi}_x + ia_{3} V \Phi \bar{\Phi}_{xx}
\right. \nn \\
&& -\; \left.  2 V D\Phi_x \bar{D}\bar{\Phi} +
2 V D\Phi \bar{D}\bar{\Phi}_x + ia_{2} V \bar{\Phi} \bar{\Phi}_{xx}
+ \frac{ia_2}{2} V\bar{\Phi}_x \bar{\Phi}_x
\right. \nonumber \\
&& +\; \left. 2 a_4 V^3 [D, \bar{D}] V +
\frac{3ia_2}{2} V^2 \Phi \Phi_x +
\frac{3a_2}{2} V [D,\bar{D}] V \Phi^2 +
    \frac{3ia_4}{2} V^2 \Phi_x \bar{\Phi}
  \right. \nonumber \\
&& - \; \left. \frac{3ia_4}{2} V^2 \Phi \bar{\Phi}_x +
3 a_{4} V [D,\bar{D}] V \Phi \bar{\Phi} -
12 DV \bar{D}V \Phi \bar{\Phi}
- \frac{3ia_{2}}{2} V^2 \bar{\Phi} \bar{\Phi}_x  \right.   \nonumber \\
&& +\; \left. \frac{3a_{2}}{2}  V [D,\bar{D}] V \bar{\Phi}^2 -
\frac{ia_{2}}{4} \bar{\Phi}_x \Phi^3 -
\frac{3ia_3}{4} \Phi^2 \bar{\Phi} \bar{\Phi}_x +
\frac{ia_2}{4} \Phi_x \bar{\Phi}^3
  \right. \nonumber \\
&& - \;\left. ia_5 V^5 - ia_2 V^3 \Phi^2
- 5i a_5 V^3 \Phi \bar{\Phi} - ia_2 V^3 \bar{\Phi}^2
- ia_{6} V\Phi^4  \right. \nn \\
&& - \; \left. \frac{ia_2}{2} V\Phi^3 \bar{\Phi} - ia_{7}
V\Phi^2 \bar{\Phi}^2 - \frac{ia_2}{2} V \Phi \bar{\Phi}^3
- ia_{6} V\bar{\Phi}^4  \right\}\;. \label{h5}
\eea
\begin{center}
\begin{tabular}{|c|c|c|c|c|c|c|c|} \hline
 & $a_1$ & $a_2$ & $a_3$ & $a_4$ & $a_5$ & $a_6$ & $a_7$ \\ \hline
a=4, b=0 & 3 & 0 & -2 & -4 & 16/5 & 0 & 6 \\ \hline
a=-2, b=6& -2 & -5 & 3 & 1 & 6/5 & 35/8 & 9/4 \\ \hline
a=-2, b=-6& -2 & 5 & 3 & 1 & 6/5 & 35/8 & 9/4 \\ \hline
\end{tabular} \vspace{0.5cm}\\
{\bf Table 2}
\end{center}

Thus, $H_5$ exists under the same restrictions \p{valuesab} on the
$N=4$ KdV
parameters $a$ and $b$ (or their manifestly $SU(2)$ covariant form
\p{consa1}, \p{consa}) as in the previous cases.
After reduction to $N=2$ super KdV by setting $\Phi = \bar \Phi = 0$,
$H_5$ is reduced to the 5 dimension conserved charges of the $a=4$ and
$a=-2$, $N=2$ KdV hierarchies, respectively,
for the first line and the last two lines in Table 2.

It is interesting to see how this conserved charge looks in the
original manifestly $N=4$ supersymmetric formulation. It is a matter
of straightforward though somewhat cumbersome computation to find that
the following $N=4$ superfield expression yields \p{h5}
after passing to $N=2$ superfields and imposing
the constraint \p{consa}
\bea
H_5 &=& {1\over 2} \int [dZ] \left[ {1\over 4} \left( D^{--}V^{++}
\right)^4 + i \left( D^{--}V^{++} \right)^2
\left( D^- \right)^2 V^{++}
\right. \nn \\
&& + \; \left. {15\over 4}\; (a^{-2})^2 \left( D^{--}V^{++} \right)^2
\left( V^{++} \right)^2
- {1\over 2} \left( D^{--}V^{++}_x \right)^2\right]  \nn \\
&& + {i\over 4} \int [d\zeta^{-2}] \left[ {63\over 100}\; (a^{-2})^4
\left( V^{++} \right)^5 - 5 \;(a^{-2})^2 \left( V^{++}_x
\right)^2 V^{++}
\right] \;. \label{harmh5}
\eea

The last conserved charge we have explicitly constructed is $H_6$.
Once again, it exists only for the above three choices of the
$N=4$ KdV
parameters. We present it here only for the choice $a=4, b=0$ since
the expressions for the two other choices are very long and
complicated. Of course, they can be obtained from the
$a=4, b=0$ expression
via finite $SU(2)$ rotations.

This charge $H_6$ reads
\begin{eqnarray}
H_6 & = & \int \mu^{(2)} \left\{ 6\; {\bar \Phi}_{xxxx} \Phi  +
12 \;V V_{xxxx}
- 240i\; {\bar D} V_x D V_x V  -
  120i\; [ D,\bar{D} ] V_{xx} V^2  \right. \nn \\
&& + \; \left. 60i\; V \bar{D} {\bar \Phi} D \Phi_{xx} +
60i\; V \bar{D} {\bar \Phi}_x D \Phi_x  +
   60i\; V \bar{D} {\bar \Phi}_{xx} D \Phi  \right. \nonumber \\
&& - \; \left. 240i\; {\bar D} V D V {\bar \Phi}_x \Phi -
  240i\; {\bar D} V D V_x {\bar \Phi} \Phi -
  60\;  [ D, \bar{D}] V [ D ,\bar{D}] V {\bar \Phi} \Phi
\right. \nonumber \\
&& -\; \left. 120\; [ D, \bar{D} ] V [ D,\bar{D}] V V^2 +
  240i\; [ D,\bar{D}] V V {\bar \Phi}_x \Phi  +
  120i\; [ D, \bar{D}] V V_x {\bar \Phi} \Phi   \right. \nonumber \\
&&+ \;\left. 120i\;[ D, \bar{D}] V_x V {\bar \Phi} \Phi  -
  15\; {\bar \Phi}^2 (\Phi_x)^2 +
  30\; {\bar \Phi} {\bar \Phi}_{xx} \Phi^2  +
  15\; ({\bar \Phi}_x)^2 \Phi^2  \right. \nonumber \\
&&+ \;\left. 240\; V^2 {\bar \Phi}_{xx} \Phi  +
  120\; V^3 V_{xx}  +
  480\; V V_x {\bar \Phi}_x \Phi  +
  360\; V V_{xx} {\bar \Phi} \Phi \right. \nn \\
&& +\;\left.  180\; (V_x)^2 {\bar \Phi} \Phi
+ 1440i\; {\bar D} V D V V {\bar \Phi} \Phi  -
   45i\; [ D, \bar{D}] V {\bar \Phi}^2 \Phi^2  \right. \nn \\
&& - \;\left.  720i\; [ D, \bar{D}] V V^2 {\bar \Phi}\Phi  -
  240i\;[ D, \bar{D}] V  V^4 +
  90\; V {\bar \Phi}^2 \Phi \Phi_x \right. \nn \\
&&- \; \left. 90\; V {\bar \Phi} {\bar \Phi}_x \Phi^2  +
 240\; V^3 {\bar \Phi} \Phi_x -
  240\; V^3 {\bar \Phi}_x \Phi  +
  20\; {\bar \Phi}^3 \Phi^3 \right. \nn \\
&& +\; \left.  360\; V^2 {\bar \Phi}^2 \Phi^2  +
480\; V^4 {\bar \Phi} \Phi +
  64i\; V^5 \right\}\;.
\end{eqnarray}

Finally, we wish to stress that all the conserved charges
$H_n, n= 1,... 6$, are in involution with respect to both
Poisson brackets
\be
\left\{ H_n, H_m \right\} = \left\{ H_n, H_m \right\}_{(1)} = 0 \;.
\ee
This property can be easily deduced from the bi-hamiltonian nature of
the conjectural $N=4$ KdV hierarchy. The bi-hamiltonian structure
can be expressed as the following general recursion relation
(up to relative scaling factors between the conserved charges)
\be
{\cal D}^{AB}\; \frac{\delta H_n}{\delta V^A} =
{\cal D}^{AB}_{(1)}\; \frac{\delta H_{n+1}}{\delta V^A} \;.
\label{bihamrec}
\ee
We have explicitly checked \p{bihamrec} for all $H_n$ presented above,
limiting ourselves, for simplicity, to the case $a=4,\; b=0$
and keeping in mind that the other two integrable cases can be generated
from this one by $SU(2)$ transformations \p{su2trans}. Actually, as
we already mentioned, {\it postulating} the relations \p{bihamrec} gives
an alternative method to construct higher-order
conservation laws, even more simple than the direct method we resorted to
in this Section. We do not foresee any reason why the construction
procedure of these laws based on the relations \p{bihamrec} should
terminate at any finite step. Both the
existence of the non-trivial conserved charges $H_2, H_4$, $H_5$
and $H_6$
and the above bi-hamiltonian property are strong indications that the
$N=4$ super KdV equation with the restrictions \p{consa1}, \p{consa}
produces the whole $N=4$ super KdV hierarchy and so is integrable.
In order to rigorously prove this, it is of primary importance to find
the appropriate Lax representation. We believe that in the $N=2$
superfield formalism this problem will be simpler than in the harmonic
superspace formulation and can be solved along the
lines of Refs. 10, 11 and 28.

\setcounter{equation}{0}
\section{Conclusion}
As the main goal of the present work, we have obtained the
$N=4$ super KdV equation of Ref. 14 in an $N=2$
superfield form and
studied the question of its integrability in this approach. We
reproduced
the results of Ref. 14 and constructed two new conserved bosonic
quantities for $N=4$ super KdV, the dimension 5 and 6 ones $H_5$ and
$H_6$. They were found to exist under the same restrictions on the
$SU(2)$ breaking parameters \p{consa1}, \p{consa} as the lower dimension
charges given in Ref. 14. The bi-hamiltonian structure of the $N=4$ KdV
equation was extended to the whole set of evolution
equation associated with the hamiltonians $H_n$ that have
been constructed. Requiring the existence of
this structure gives rise to the same conditions \p{consa1},
\p{consa} on the
parameters. These results suggest that the unique integrable
$N=4$ $SU(2)$ KdV hierarchy exists, with the choice
of the $SU(2)$ breaking parameters as in Eqs. \p{consa1}, \p{consa}.
The $N=2$ superfield formulation allowed us also to show that
two inequivalent reductions to $N=2$ KdV are possible. They yield,
respectively, the integrable $a=4$ and $a=-2$ cases of $N=2$ KdV. Thus
the single $N=4$ $SU(2)$ KdV hierarchy incorporates as particular
solutions two of the three $N=2$ KdV hierarchies.

Among the problems for future study, besides the construction of
a Lax pair representation for the $N=4$ $SU(2)$ KdV, let us mention a
generalization to the case of the ``large''
$N=4$ superconformal algebra \cite{{belg},{IKL}} with the
affine subalgebra $so(4) \times u(1)$. The related $N=4$ super KdV
hierarchy is expected to embrace both the $N=4$ $SU(2)$ and $N=3$ KdV
ones as particular cases.
Also, it would be interesting to construct generalized $N=4$ super
KdV systems associated with
nonlinear $W$ type extensions of $N=4$ superconformal algebras.
One of possible ways to define such extensions was mentioned
in Subsec. 2.1.

\vspace{0.5cm}

\noindent{\Large\bf Acknowledgements}

\vspace{0.3cm}
\noindent E.I. is grateful to P. Mathieu for useful discussions.
He also thanks ENSLAPP, ENS-Lyon, for hospitality extended to him
during the course of this work. E.I. and S.K. thank the Russian
Foundation of Fundamental Research, grant 93-02-03821, and the
International Science Foundation, grant M9T300, for financial support.

\setcounter{equation}{0}
\vspace{0.7cm}

\noindent{\Large\bf Appendix A}
\def\theequation{A.\arabic{equation}}

\vspace{0.3cm}
\noindent In this Appendix we collect a number of useful identities.

First of all, we present some consequences of the constraints
\p{ct1} and their harmonic superspace version \p{ct2}, \p{ct3}:
\bea
D^i V^{kl} &=& -{1\over 3} \left( \epsilon^{ik} \xi^l +
\epsilon^{il} \xi^k \right)\;,
\;\;
\bar D^i V^{kl} \;=\; {1\over 3} \left( \epsilon^{ik} \bar \xi^l +
\epsilon^{il} \bar \xi^k \right)\;,  \\
D^i \bar D^j V^{kl} &=& -{i\over 2} \left( \epsilon^{jk} V^{il}_x
+ \epsilon^{jl} V^{ik}_x \right) -{1\over 6} \left( \epsilon^{il}
\epsilon^{jk} + \epsilon^{ik} \epsilon^{jl} \right) T \\
D^i D^j V^{kl} &=& \bar D^i \bar D^j V^{kl} \;=\; 0\;, \\
D^{-}V^{++} &=& -{2\over 3} \xi^k u^+_k\;, \;\;
\bar D^{-}V^{++} = {2\over 3} \bar \xi^k u^+_k\;, \\
(D^-)^2 V^{++} &=& -{i\over 2}\;D^{--} V^{++}_x -{1\over 3} \;T\;.
\eea
\vspace{0.3cm}

When deducing Eqs. \p{kdv4} and \p{kdv2} from the harmonic superspace
Poisson structure \p{poi} and rewriting the latter in ordinary
$N=4$ superspace, one needs to decompose the objects given in terms
of one set of harmonic variables, say $u^{\pm}_i$, over another set,
$v^{\pm}_i$, using the completeness condition \p{ct1c}. The general
decomposition formula for some object bilinear in harmonics,
$$
S^{++}(u) \equiv S^{ik} u^+_iu^+_k
$$
($S^{++}$ can stand, e.g., for $V^{++}$ or $(D^+)^2 = D^+\bar D^+$),
is as follows
\be
S^{++}(u) = S^{++}(v) (v^-u^+)^2 + {1\over 2} (D^{--}_v)^2 S^{++}(v)
(v^+u^+)^2 - D^{--}_v S^{++}(v) (v^- u^+)(v^+u^+)\;. \label{decs}
\ee
Analogous relations for other harmonic projections of $S^{ik}$,
namely $S^{+-}$ and $S^{--}$, can be obtained by applying
$D^{--}_u$ to both sides of Eq. \p{decs} and making use of the
harmonic differentiation rules
$$
D^{--} u^+_i = u^-_i\;, \;\; D^{--} u^-_i  = 0\;.
$$

\setcounter{equation}0
\vspace{0.7cm}

\noindent{\Large\bf Appendix B}
\def\theequation{B.\arabic{equation}}

\vspace{0.3cm}
\noindent In this Appendix we prove the following Lemma.

{\it Lemma}: Let $c^{iklj}$ be an arbitrary rank 4 symmetric
$SU(2)$ spinor subjected to the reality condition
$$
(c^{iklj})^\dagger = \epsilon_{ii'}\epsilon_{kk'}\epsilon_{ll'}
\epsilon_{jj'}
c^{i'k'l'j'}.
$$
The
necessary and sufficient conditions for it to be a square of some real
rank 2 symmetric $SU(2)$ spinor $a^{ik}$,
\be
c^{ijkl} = {1\over 3} \left( a^{ij} a^{kl} + a^{ik}a^{jl} + a^{il}a^{jk}
\right)\;,  \;\;\; (a^{ik})^\dagger = -\epsilon_{il} \epsilon_{kj}
a^{lj}\;,
\label{usllem}
\ee
are the following ones
\be
(I) \;{\cal A}^3 = 6\; {\cal B}^2\;, \;\;(II)\;{\cal B} < 0\;; \;\;\;
\left( {\cal A} \equiv c^{ijkl}\;c_{ijkl}, \;\; {\cal B}
\equiv c^{ik}_{\;\;\;jl}\;
c^{jl}_{\;\;\;ft}\;
c^{ft}_{\;\;\;ik}\;\right). \label{ABcond}
\ee

{\it Proof}: The proof is simpler in the vector notation,
with $c^{ikjl}$ represented
by a real traceless symmetric rank 2 tensor and $a^{ik}$ by a
real vector
$$
c^{iklj} \Rightarrow c^{\mu \nu} = {1\over 2} c^{ij}_{\;\;\;kl}
(\sigma^\mu)_{i}^k
(\sigma^\nu)_{j}^l\;,\;\;
a^{ik} \Rightarrow a^\mu =
{1\over \sqrt{2}} a^{i}_{\;k} (\sigma^\mu)_{i}^k\;; \;\;
(\mu, \nu ... = 1,2,3)
$$
$$
{\cal A} = c^{\mu \nu} c^{\mu \nu}\;, \;\;\;
{\cal B} = - c^{\mu \nu} c^{\nu \rho} c^{\rho \mu}
\;.
$$
Here, $(\sigma^\mu)^l_k$ are Pauli matrices.

In this notation, the relation \p{usllem} amounts to
\be
c^{\mu \nu} = a^\mu a^\nu - {1\over 3}\;\delta^{\mu\nu} \;
(a^\rho a^\rho)\;.
\label{vectc}
\ee
Then the necessity of \p{ABcond} immediately follows from computing
the invariants ${\cal A}$ and ${\cal B}$ for the tensor \p{vectc}
$$
{\cal A} = {2\over 3}\; (a^2)^2\;, \;\; {\cal B} =
-{2\over 9}\; (a^2)^3\;, \;\; a^2 \equiv
a^\mu a^\mu > 0\;.
$$

In order to show that \p{ABcond} is also sufficient for
$c^{\mu \nu}$ to
be representable in the form \p{vectc}, let us go to the frame where
$c^{\mu\nu}$ is a diagonal traceless matrix with the following non-zero
entries
\be
c^{11} = \lambda_1,\;\; c^{22} = \lambda_2, \;\;
c^{33} = - (\lambda_1 +\lambda_2)\;,
\label{partic}
\ee
$\lambda_1$, $\lambda_2$ being arbitrary for the moment.
After substituting this into the first of conditions \p{ABcond}
we get the equation
\be
(\lambda_1 - \lambda_2)^2\;
(\lambda_1 + 2\lambda_2)\;(2 \lambda_1 + \lambda_2) = 0 \;,
\ee
which has the following non-zero roots
\be
(a)\; \lambda_1 = \lambda_2;\; \;\;(b)\; \lambda_1 = -2 \lambda_2;\;\;\;
(c)\; \lambda_2 = - 2 \lambda_1 \;.
\label{roots}
\ee
The inequality in \p{ABcond} takes the form
\be
\lambda_1\lambda_2 \;(\lambda_1 + \lambda_2) < 0
\ee
and restricts the solutions \p{roots} in the following way
\be
(a)\; \lambda_1 < 0\;, \;\;\;(b)\; \lambda_1 > 0\;, \;\;\;
(c)\; \lambda_1 < 0 \;.
\ee
Now it is an elementary exercise to see that these three solutions
correspond to three different choices of the vector $a^\mu$ in \p{vectc}
(up to the reflection $a^\mu \rightarrow - a^\mu $)
$$
(a)\; a^\mu = (0, 0, \sqrt{3|\lambda_1|})\;; \;
(b) \;a^\mu = (\sqrt{{3\over 2} |\lambda_1|}, \;0, \;0)\;; \;
(c) \;a^\mu = (0, \; \sqrt{3|\lambda_1|}, \;0)\;.
$$
This proves the sufficiency of the conditions (B.2).

\end{document}